\documentclass[ALICE,manyauthors]{cernphprep}

\usepackage[comma,square,numbers,sort&compress]{natbib}
\usepackage{hyperref}
\usepackage{lineno}
\usepackage{multirow}
\usepackage{amsmath}

\newcommand{\La}{\ensuremath{\Lambda}\,}
\newcommand{\aLa}{\ensuremath{\overline{\Lambda}}\,}
\newcommand{\pp}{\ensuremath{\mathrm {p\kern-0.05em p}}\,}

\newcommand{\pP}{\ensuremath{\mbox{p--p}}\,}

\newcommand{\pL}{\ensuremath{\mbox{p--$\Lambda$}}\,}

\newcommand{\LL}{\ensuremath{\mbox{$\Lambda$--$\Lambda$}}\,}
\newcommand{\nL}{\ensuremath{\mbox{N--$\Lambda$}}\,}

\newcommand{\pSiZero}{\ensuremath{\mbox{p--$\Sigma^0$}}\,}
\newcommand{\pXim}{\ensuremath{\mbox{p--$\Xi$}}\,}
\newcommand{\pXi}{\ensuremath{\mbox{p--$\Xi$}}\,}
\newcommand{\nXi}{\ensuremath{\mbox{N--$\Xi$}}\,}

\newcommand{\radiusResult}{\ensuremath{r_0 = 1.125 \pm 0.018\, \mathrm{(stat)} \,^{+0.058} _{-0.035}\, \mathrm{(syst)}}\,fm}

\newcommand{\MeVc}{\ensuremath{\mathrm{MeV}\kern-0.05em/\kern-0.02em c\,}}
\newcommand{\GeVc}{\ensuremath{\mathrm{GeV}\kern-0.05em/\kern-0.02em c \,}}
\newcommand{\GeVcSq}{\ensuremath{\mathrm{GeV}\kern-0.05em/\kern-0.02em c^2\,}}
\newcommand{\MeVcSq}{\ensuremath{\mathrm{MeV}\kern-0.05em/\kern-0.02em c^2\,}}

\newcolumntype{b}{X}
\newcolumntype{s}{>{\hsize=.5\hsize}X}
\newcommand{\heading}[1]{\multicolumn{1}{c}{#1}}

\begin{document}%

\begin{titlepage}
\PHyear{2018}
\PHnumber{150}      
\PHdate{30 May}  
%

\title{p$\mbox{--}$p, p$\mbox{--}\Lambda$ and $\Lambda\mbox{--}\Lambda$ correlations studied via femtoscopy in p\kern-0.05em p reactions at $\mathbf{\sqrt{s}=7}$\,TeV}
\ShortTitle{p$\mbox{--}$p, p$\mbox{--}\Lambda$  and $\Lambda\mbox{--}\Lambda$ correlations studied via femtoscopy in p\kern-0.05em p at $\sqrt{s}=7$\,TeV}

\Collaboration{ALICE Collaboration\thanks{See Appendix~\ref{app:collab} for the list of collaboration members}}
\ShortAuthor{ALICE Collaboration} 

\begin{abstract}
We report on the first femtoscopic measurement of baryon pairs, such as \pP,
\pL and \LL, measured by ALICE at the Large Hadron Collider (LHC) in proton-proton collisions at $\sqrt{s} = 7$\,TeV. This study demonstrates the
feasibility of such measurements in \pp collisions at ultrarelativistic
energies. The femtoscopy method is employed to constrain the hyperon\mbox{--}nucleon and hyperon\mbox{--}hyperon interactions, which are still rather poorly understood.
A new method to evaluate the influence of residual correlations
induced by the decays of resonances and experimental impurities is hereby
presented. 
The \pP, \pL and \LL correlation functions were fitted simultaneously with the help of a new tool developed specifically for the femtoscopy analysis 
in small colliding systems ``Correlation Analysis Tool using the
Schr\"{o}dinger Equation'' (CATS). Within the assumption that in \pp collisions the three particle pairs originate from a common source, its radius is found to be equal to \radiusResult. The sensitivity
of the measured \pL correlation is tested against different scattering
parameters which are defined by the interaction among the two particles, but the statistics is not sufficient yet to discriminate among
different models. The measurement of the \LL correlation function constrains
the phase space spanned by the effective range and scattering length of the
strong interaction. Discrepancies between the measured scattering parameters and the resulting correlation functions at LHC and RHIC energies are discussed in the context of various models.
\end{abstract}
\end{titlepage}
\setcounter{page}{2}

\section{Introduction}
Traditionally femtoscopy is used in heavy-ion collisions at ultrarelativistic
energies to investigate the spatial-temporal evolution of the particle emitting
source created during the collision~\cite{Pratt:1986cc,Lisa:2005dd}. 
Assuming that the interaction for the employed particles is known, a
detailed study of the geometrical extension of the emission region becomes
possible~\cite{Henzl:2011dh,Agakishiev:2011zz,Kotte:2004yv,Aggarwal:2007aa,Adams:2004yc,Aamodt:2011mr,Abelev:2014pja,Adam:2015vna}.\\
If one considers smaller colliding systems such as proton-proton ($\mathrm{p\kern-0.05em p}$) at TeV
energies and assumes that the particle emitting source does
not show a strong time dependence, one can reverse the paradigm and exploit
femtoscopy to study the final state interaction (FSI). 
This is especially interesting in the case where the interaction strength is 
not well known as for hyperon\mbox{--}nucleon (Y\mbox{--}N) and hyperon\mbox{--}hyperon
(Y\mbox{--}Y) pairs~\cite{PhysRevLett.54.302,Adams:2005ws,Anticic:2011ja,Chung:2002vk,Agakishiev:2010qe,Adamczyk:2014vca,Adamczyk:2015hza,Shapoval:2014yha,Kisiel:2014mma}. \\
Hyperon\mbox{--}nucleon and hyperon\mbox{--}hyperon interactions are still rather poorly
experimentally constrained and a detailed knowledge of these interactions is 
necessary to understand quantitatively the strangeness sector in the low-energy
regime of Quantum-Chromodynamics (QCD)~\cite{Weise:2009ny}.\\
Hyperon\mbox{--}nucleon (\pL and p$\mbox{--}\Sigma$) scattering experiments have been
carried out in the sixties~\cite{SechiZorn:1969hk,Eisele:1971mk,Alexander:1969cx} 
and the measured cross sections have been used to extract scattering lengths
and effective ranges for the strong nuclear potential by means of effective
models such as the Extended-Soft-Core (ESC08) baryon\mbox{--}baryon model
\cite{Nagels:2015lfa} or by means of chiral effective field theory ($\chi$EFT) approaches
at leading order (LO)~\cite{POLINDER2006244} and next-to-leading order (NLO)
\cite{Haidenbauer:2013oca}.
The results obtained from the above-mentioned models are rather different, but
all confirm the attractiveness of the \La\mbox{--}nucleon ($\Lambda$\mbox{--}N) interaction for low
hyperon momenta. In contrast to the LO results, the NLO solution claims the
presence of a negative phase shift  in the \pL spin singlet channel for \La
momenta larger than $p_{\Lambda}>600$\,MeV/$c$. This translates into a
repulsive core for the strong interaction evident at small relative distances.
The same repulsive interaction is obtained in the p-wave channel within the
ESC08 model~\cite{Nagels:2015lfa}.\\
The existence of hypernuclei~\cite{Hashimoto:2006aw} confirms that the \nL is
attractive within nuclear matter  for densities below nuclear saturation
$\rho_0=\,0.16 ~\mathrm{fm}^{-3}$. An average value of
$U(\rho=\rho_0,k=0)\approx\,-30$\,MeV \cite{Hashimoto:2006aw}, with $k$ the hyperon momentum in the laboratory
reference system, is extracted from  hypernuclear data
 on the basis of a dispersion relation for hyperons in a
baryonic medium at $\rho_0$ . \\
The situation for the $\Sigma$ hyperon is currently rather unclear. There 
are some experimental indications for the formation of $\Sigma$\mbox{--}hypernuclei \cite{Hayano:1988pn,PhysRevLett.80.1605} but
 different theoretical
approaches predict both attractive and repulsive interactions depending on the
isospin state and partial
wave~\cite{Nemura:2017vjc,Nagels:2015lfa,Haidenbauer:2013oca}. The scarce experimental 
data for this hypernucleus prevents any validation of
the models.\\
A $\Xi$\mbox{--}hypernucleus candidate was detected~\cite{Nakazawa:15} and ongoing
measurements suggest that the \nXi inter-action is weakly
attractive~\cite{Nagae:2017slp}. A recent work by the Lattice HAL-QCD
Collaboration~\cite{Hatsuda:2017uxk} shows how this attractive interaction could be visible in the
\pXi femtoscopy analysis, in particular by comparing correlation functions for
different static source sizes. This further motivates
the extension of the femtoscopic studies from heavy ions to \pp collisions since
in the latter case the source size decreases by about a factor of three at the
LHC energies leading to an increase in the strength of the correlation
signal~\cite{PhysRevLett.83.3138}.   \\
If one considers hyperon\mbox{--}hyperon interactions, the most prominent example is
the \LL case. The H-dibaryon \LL bound state was
predicted~\cite{PhysRevLett.38.195} and later  a double $\Lambda$ hypernucleus
was observed ~\cite{Takahashi:2001nm}. From this single measurement a shallow
\LL binding energy of few MeV was extracted, but the H-dibaryon state was never
observed. Also recent lattice calculations~\cite{Sasaki:2017ysy} obtain a
rather shallow attraction for the \LL state. \\
The femtoscopy technique was employed by the STAR collaboration to
study \LL correlations in Au\mbox{--}Au collisions at $\sqrt{s_{\mathrm{NN}}} =
200$\,GeV~\cite{Adamczyk:2014vca}. First a shallow repulsive interaction was
reported for the \LL system, but in an alternative analysis, where the residual
correlations were treated more accurately~\cite{PhysRevC.91.024916}, a shallow
attractive interaction was confirmed. These analyses demonstrate the 
limitations of such measurements in heavy-ion collisions, where the source
parameters are time-dependent and the emission time might not be the same
for all hadron species.\\
The need for more experimental data to study the hyperon\mbox{--}nucleon,
hyperon\mbox{--}hyperon and even the hyperon\mbox{--}nucleon\mbox{--}nucleon interaction has become
more crucial in recent years due to its connection to the modeling of
astrophysical objects like neutron stars 
~\cite{Petschauer:2015nea,Schulze:2006vw,Weissenborn:2011kb,Weissenborn:2011ut}. 
In the inner core of these objects the appearance of hyperons is a possible
scenario since their creation at finite density becomes energetically favored
in comparison with a purely neutron matter composition
~\cite{Weissenborn:2011kb}. However, the appearance of these additional degrees
of freedom leads to a softening of the nuclear matter equation of state
(EOS)~\cite{Djapo:2008au} making the EOS incompatible with the observation of 
neutron stars as heavy as two solar masses~\cite{Demorest:2010bx, Antoniadis:2013pzd}.
This goes under the name of the 'hyperon puzzle'.  Many attempts were made to
solve this puzzle, e.g. by introducing three-body forces for $\Lambda$NN
leading to an additional repulsion that can counterbalance the large
gravitational pressure and finally allow for larger neutron star
masses~\cite{Yamamoto:2013ada,Yamamoto:2014jga,Oertel:2016bki,Lonardoni:2014bwa}.
A repulsive core for the two body forces would also stiffen the EOS containing
hyperons. \\
In order to constrain the parameter space of such models a detailed knowledge
of the hyperon\mbox{--}nucleon, including $\Sigma$ and $\Xi$ states, and of the
hyperon\mbox{--}nucleon\mbox{--}nucleon interaction is mandatory.\\
This work presents an alternative to scattering experiments, using the
femtoscopy technique to study the Y\mbox{--}N and Y\mbox{--}Y interactions
in \pp collisions at $\sqrt{s}=7\,$TeV.  We show that \pp collisions at the LHC
are extremely well suited to investigate baryon\mbox{--}baryon final state interactions
and that the measurement of the correlation function is not contaminated with
the mini-jet background visible in meson\mbox{--}meson 
correlations~\cite{PhysRevD.84.112004,Abelev:2012sq}.
The extracted \pP, \pL and \LL correlations have been compared to the predicted
function obtained by solving the Schr\"odinger equation exactly by employing
the Argonne $v_{18}$ potential \cite{PhysRevC.51.38} for \pP pairs and different scattering parameters
available in the literature for \pL and \LL pairs. The predictions for the
correlation function used to fit the data are obtained with the newly developed
CATS framework~\cite{CATS}. A common source with a constant size is assumed and
the value of the radius is extracted.  \\
The work is organized in the following way: in Section II the experiment setup
and the analysis technique are briefly introduced. In Section III the femtoscopy
technique and the theoretical models employed are discussed. In Section IV the
sources of systematic uncertainties are summarized and finally in Section V the
results for the \pP, \pL and \LL correlation function are presented.

\section{Data analysis}
In this paper we present results from studies of the \pP, \pL and \LL
correlations in \pp collisions at $\sqrt{s}=7$\,TeV employing the data collected by ALICE
in 2010 during the LHC Run 1. Approximately $3.4 \times 10^{8}$ minimum bias events have been used
for the analysis, before event and track selection. A detailed description of
the ALICE detector and its performance in the LHC Run 1 (2009-2013) is given in
\cite{1748-0221-3-08-S08002, Abelev:2014ffa}. The inner tracking system (ITS)
\cite{1748-0221-3-08-S08002} consists of six cylindrical layers of high
resolution silicon detectors placed radially between $3.9$ and $43$\,cm around
the beam pipe. The two innermost layers are silicon pixel detectors (SPD) and
cover the pseudorapidity range $|\eta|<2$. The time projection chamber (TPC)
\cite{2010TPCNIMA} provides full azimuthal coverage and allows charged particle
reconstruction and identification (PID) via the measurement of the specific
ionization energy loss d$E$/d$x$ in the pseudorapidity range $|\eta| < 0.9$.
The Time-Of-Flight (TOF)~\cite{Akindinov2013} detector consists of Multigap
Resistive Plate Chambers covering the full azimuthal angle in $|\eta|<0.9$. The
PID is obtained by measuring the particle's velocity $\beta$. The above
mentioned detectors are immersed in a $B = 0.5$\,T solenoidal magnetic field
directed along the beam axis. The V0 are small-angle plastic scintillator
detectors used for triggering and placed on either side of the collision vertex
along the beam line at $+3.3$\,m and  $-0.9$\,m from the nominal interaction
point, covering the pseudorapidity ranges $2.8 < \eta < 5.1$ (V0-A) and $-3.7 <
\eta < -1.7$ (V0-C).

\subsection{Event selection}
The minimum bias interaction trigger requires at least two out of the following
three conditions: two pixel chips hit in the outer layer of the silicon pixel
detectors, a signal in V0-A, a signal in V0-C~\cite{Abelev:2014ffa}. Reconstructed events are
required to have at least two associated tracks and the distance along the beam
axis between the reconstructed primary vertex and the nominal interaction point
should be smaller than 10\,cm. 
Events with multiple reconstructed SPD vertices are considered as pile-up. In addition, background events are rejected using the correlation between the number of SPD clusters and the tracklet multiplicity. 
The tracklets are constrained to the primary vertex, and hence a typical background event is characterized by a large amount of SPD clusters but only few tracklets, while a pile-up event contains a larger number of clusters at the same tracklet multiplicity.

After application
of these selection criteria, about $2.5\times10^{8}$ million events are
available for the analysis.
 
\subsection{Proton candidate selection}
To ensure a high purity sample of protons, strict selection criteria are imposed
on the tracks. Only particle tracks reconstructed with the TPC without
additional matching with hits in the ITS are considered in
the analysis in order to avoid biases introduced by the non-uniform acceptance in the ITS. 
However, the track fitting is constrained by the
independently reconstructed primary vertex. Hence, the obtained momentum
resolution is comparable to that of globally reconstructed tracks, as
demonstrated in~\cite{Abelev:2014ffa}.

The selection criteria for the proton candidates are summarized in
Tab.~\ref{tab:Cuts}. The selection on the number of reconstructed TPC clusters
serves to ensure the quality of the track, to assure a good $p_{\mathrm{T}}$
resolution at large momenta and to remove fake tracks from the sample.
To enhance the number of protons produced at the primary vertex, a selection is imposed on the distance-of-closest-approach (DCA) in both beam ($z$) and transverse ($xy$) directions.
In order to minimize the fraction of protons originating from the interaction of
primary particles with the detector material, a low transverse momentum cutoff
is applied~\cite{Adam2017}. At high $p_{\mathrm{T}}$ a cutoff is introduced to
ensure the purity of the proton sample, as the purity drops below 80\,\% for
larger $p_{\mathrm{T}}$ due to the decreasing separation power of the combined
TPC and TOF particle identification.

For particle identification both the TPC and the TOF detectors are employed.
For low momenta ($p < 0.75$\,GeV/$c$) only the PID selection from the TPC is
applied, while for larger momenta the information of both detectors is combined
since the TPC does not provide a sufficient separation power in this momentum
region. The combination of TPC and TOF signals is done by employing a circular
selection criterion 
$n_{\sigma,\mathrm{combined}}\equiv\sqrt{(n_{\sigma,\mathrm{TPC}})^2+(n_{\sigma,\mathrm{TOF}})^2}$, 
where $n_{\sigma}$ is the number of standard deviations of the measured from
the expected signal at a given momentum. The expected signal is computed in the
case of the TPC from a parametrized Bethe\mbox{--}Bloch curve, and in the case of the
TOF by the expected $\beta$ of a particle with a mass hypothesis $m$. In
order to further enhance the purity of the proton sample, the $n_{\sigma}$ is
computed assuming different particle hypotheses (kaons, electrons and pions)
and if the corresponding hypothesis is found to be more favorable, i.e. the
$n_{\sigma}$ value found to be smaller, the proton hypothesis and thus the
track is rejected.
With these selection criteria a $p_{\mathrm{T}}$-averaged proton purity of 99\,\% is achieved. The purity remains above 99\,\% for $p_{\mathrm{T}} < 2$\,GeV/\textit{c} and then decreases to 80\,\% at the momentum cutoff of 4.05\,GeV/\textit{c}.

\begin{table}[t]
\begin{center}    
\begin{tabular}{l | l}
\hline  \hline
Selection criterion & Value \\
\hline 
\textbf{Proton selection criteria} & \\
Pseudorapidity & $|\eta|< 0.8$  \\
Transverse momentum & $0.5 < p_{\mathrm{T}} < 4.05$\,GeV/$c$ \\
TPC clusters & $n_{\mathrm{TPC}}> 80$ \\
Crossed TPC pad rows & $n_{\mathrm{crossed}}> 70$ (out of 159) \\
Findable TPC clusters & $n_{\mathrm{crossed}}/n_{\mathrm{findable}} > 0.83$ \\
Tracks with shared TPC clusters & rejected \\
Distance of closest approach $xy$ & $|\mathrm{DCA}_{xy}| < 0.1$\,cm \\
Distance of closest approach $z$ & $|\mathrm{DCA}_{z}| < 0.2$\,cm \\
\multirow{2}{*}{Particle identification} & $|n_{\sigma,\mathrm{TPC}}| < 3$ for $p < 0.75$\,GeV/$c$  \\
  & $n_{\sigma,\mathrm{combined}}<3$ for $p > 0.75$\,GeV/$c$ \\
\hline 
\textbf{Lambda selection criteria} & \\
\textit{Daughter track selection criteria} & \\
Pseudorapidity & $|\eta|< 0.8$  \\
TPC clusters & $n_{\mathrm{TPC}}> 70$ \\
Distance of closest approach  & $\mathrm{DCA} > 0.05$\,cm \\
Particle identification & $|n_{\sigma,\mathrm{TPC}}|<5$ \\
& \\
 $V_0$ \textit{selection criteria} & \\
 Transverse momentum & $p_{\mathrm{T}} > 0.3$\,GeV/$c$ \\
 \La decay vertex & $|i_{\mathrm{vertex_{\Lambda}}}| < 100$\,cm, $i$=$x$,$y$,$z$ \\
 Transverse radius of the decay vertex $r_{xy}$ & 0.2$ < r_{xy} < $100\,cm \\
DCA of the daughter tracks at the decay vertex & $\mathrm{DCA}(|p,\pi|) < $1.5\,cm \\
 Pointing angle $\alpha$ & $\cos \alpha > 0.99$ \\
K$^{0}$ rejection &  $0.48 < \mathrm{M}_{\pi^+ \pi^-} < 0.515$\,GeV/$c^{2}$ \\
\La selection & $|\mathrm{M}_{\mathrm{p}\pi} - \mathrm{M}_{\Lambda,\mathrm{PDG}}|<4$\,MeV/$c^{2}$ \\
\hline \hline
\end{tabular}
\caption[Proton cuts]{Proton (\textit{top}) and \La candidate (\textit{bottom}) selection criteria.}
\label{tab:Cuts}
\end{center}
\end{table}

\subsection{$\mathbf{Lambda}$ candidate selection}
The weak decay $\Lambda\rightarrow \mathrm{p}\pi^-$ (BR= 63.9\,\%, $c\tau = 7.3$\,cm~\cite{Patrignani:2016xqp}) is exploited for the reconstruction of the $\Lambda$ candidate, and accordingly the charge-conjugate decay for the \aLa identification. The reconstruction method forms so-called $V_0$ decay candidates from two charged particle tracks using a procedure described in~\cite{0954-3899-32-10-001}.
The selection criteria for the \La candidates are summarized in Tab.~\ref{tab:Cuts}. 
The $V_0$ daughter tracks are globally reconstructed tracks and, in order to maximize the efficiency, selected by a broad particle identification cut employing the TPC information only. Additionally, the daughter tracks are selected by requiring a minimum impact parameter of the tracks with respect to the primary vertex. After the selection all positively charged daughter tracks are combined with a negatively charged partner to form a pair. 
The resulting \La vertex $i_{\mathrm{vertex_{\Lambda}}}$, $i$=$x$,$y$,$z$ is then defined as the point of closest approach between the two daughter tracks. This distance of closest approach of the two daughter tracks with respect to the \La decay vertex $\mathrm{DCA}(|p,\pi|)$ is used as an additional quality criterion of the \La candidate.

The \La momentum is calculated as the sum of the daughter momenta. A minimum transverse momentum requirement on the \La candidate is applied to reduce the contribution of fake candidates.
Finally, a selection is applied on the opening angle $\alpha$ between the \La momentum and the vector pointing from the primary vertex to the secondary $V_0$ decay vertex.
The rather broad PID selection of the daughter tracks introduces a residual pion contamination of the proton daughter sample that in combination with the charge-conjugate pion of the $V_0$ leads to the misidentification of $K^0_S$ as \La candidates. These $K^0_S$ candidates are removed by a selection on the $\pi^+\pi^-$ invariant mass.

\begin{figure}
\centering
\includegraphics[width=.85\textwidth]{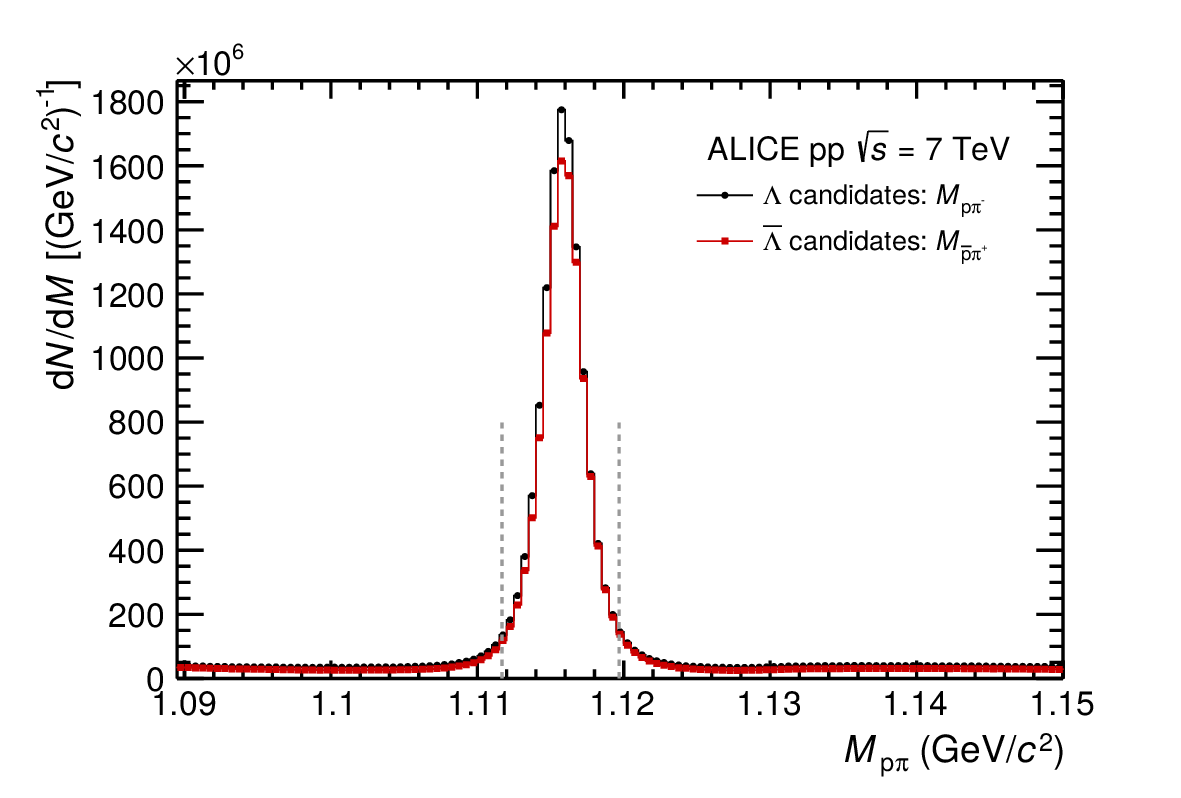}
\caption{\textit{(Color online)} Invariant mass distribution of $\mathrm{p}\pi^-$ ($\overline{\mathrm{p}}\pi^+$) to obtain the \La (\aLa) signal. The
dashed lines set the selection width used in the analysis.}
\label{fig:V0signal}
\end{figure}

The reconstructed invariant mass, its resolution and purity are determined by fitting eight spectra of the same size in $p_{\mathrm{T}} \in [0.3, ~4.3]$\,GeV/$c$ with the sum of two Gaussian functions describing the signal and a second-order polynomial to emulate the combinatorial background. The obtained values for the mean and variance of the two Gaussian functions are combined with an arithmetic average. The determined mass is in agreement with the PDG value for the \La and \aLa particles~\cite{Patrignani:2016xqp}.
A total statistics of $5.9\times10^{6}$ and $5.5\times10^{6}$ and a signal to background ratio of 20 and 25 at a $p_{\mathrm{T}}$-averaged purity of 96\,\% and 97\,\% is obtained for \La and \aLa, respectively. It should be noted that the \La purity is constant within the investigated $p_{\mathrm{T}}$ range.
Finally, a selection on the $p\pi^-$ ($\overline{\mathrm{p}}\pi^+$) invariant mass is applied. 
To avoid any contribution from auto-correlations, all \La candidates are checked for shared daughter tracks. If this condition is found to be true, the \La candidate with the smaller cosine pointing angle is removed from the sample. If a primary proton is also used as a daughter track of a \La candidate, the latter is rejected. 
Figure~\ref{fig:V0signal} shows the $p_{\mathrm{T}}$-integrated invariant mass of the \La and \aLa candidates. 

\section{The correlation function}
 \label{Sec:CF}
The observable of interest in femtoscopy is the two-particle correlation
function, which is defined as the probability to find simultaneously two
particles with momenta $\mathbf{p}_1$ and $\mathbf{p}_2$ divided by the product
of the corresponding single particle probabilities
\begin{equation}
C(\mathbf{p}_1,\mathbf{p}_2)\equiv
\frac{P(\mathbf{p}_1,\mathbf{p}_2)}{P(\mathbf{p}_1)\cdot P(\mathbf{p}_2)}.
\label{eq:CFtheo}
\end{equation}
These probabilities are directly related to the inclusive Lorentz invariant
spectra $P(\mathbf{p}_1,\mathbf{p}_2)=E_1 E_2
\frac{\mathrm{d}^6N}{\mathrm{d}^3p_1 \mathrm{d}^3p_2}$ and
$P(\mathbf{p}_{1,2})=E_{1,2}\frac{\mathrm{d}^3N}{\mathrm{d}^3p_{1,2}}$.
In absence of a correlation signal the value of $C(\mathbf{p}_1,\mathbf{p}_2)$
equals unity.

Approximating the emission process and the momenta of the particles, the size
of the particle emitting source can be studied. Following~\cite{Lisa:2005dd},
Eq. (\ref{eq:CFtheo}) can then be rewritten as
\begin{equation}
C(\mathbf{k^*})= \int \, d^3 r^* S(r^*) |\psi(r^*,\mathbf{k^*})|^2,
\label{eq:CFsourcewf}
\end{equation}
where $k^*$ is the relative momentum of the pair defined as
$k^*=\frac{1}{2}\cdot|\mathbf{p}^*_1-\mathbf{p}^*_2|$, with $\mathbf{p}^*_1$
and $\mathbf{p}^*_2$ the momenta of the two particles in the pair rest frame (PRF, denoted by the $^*$), 
$S(r^*)$ contains the distribution of the relative distance of particle pairs
in the pair rest frame, the so-called source
function, and $\psi(r^*,\mathbf{k^*})$ denotes the relative wave function of
the particle pair. The latter contains the particle interaction term and determines
the shape of the correlation function. In this work, the \pP correlation
function, which is theoretically well understood, is employed to obtain the
required information about the source function and this information will be
used to study the \pL and \LL interaction.

In order to relate the correlation function to experimentally accessible
quantities, Eq. (\ref{eq:CFtheo}) is reformulated~\cite{Lisa:2005dd} as 
\begin{equation}
C(k^*)=\mathcal{N}\frac{A(k^*)}{B(k^*)},
\label{eq:CFexp}
\end{equation}
The distribution of particle pairs from the same event is denoted with $A(k^*)$
and $B(k^*)$ is a reference sample of uncorrelated pairs. The latter is
obtained using event mixing techniques, in which the particle pairs of interest
are combined from single particles from different events. To avoid
acceptance effects of the detector system, the mixing procedure is conducted
only between particle pairs stemming from events with similar $z$ position of
the primary vertex and similar multiplicity~\cite{Lisa:2005dd}. 
The normalization parameter for mixed and same event yields $\mathcal{N}$ is chosen
such that the mean value of the correlation function equals unity for $k^*\in
[0.2, ~0.4]$\,GeV/$c$.
 
As correlation functions of all studied baryon\mbox{--}baryon pairs, i.e. \pP,
\pL and \LL, exhibit identical behavior compared to those of their respective
anti-baryon\mbox{--}anti-baryon pairs, the corresponding samples are combined
to enhance the statistical significance. Therefore, in the following \pP
denotes the combination of $\mbox{p--p} \oplus
\overline{\mathrm{p}}\mbox{--}\overline{\mathrm{p}}$, and accordingly for \pL
and \LL.

\subsection{Decomposition of the correlation function} 
\label{deco}

The experimental determination of the correlation function is distorted by two
distinct mechanisms. The sample of genuine particle pairs
include misidentified particles and feed-down particles from strong and weak
decays.

In this work a new method to separate all the individual components
contributing to a measured correlation signal is proposed. The correlation
functions arising from resonances or impurities of the sample are weighted
with the so-called $\lambda$ parameters and in this way are taken into account
in the total correlation function of interest
\begin{equation}
C(k^*) = 1 + \lambda_{\mathrm{genuine}} \cdot ( C_{\mathrm{genuine}}(k^*)-1 )
+\sum_{ij}\lambda_{ij}(C_{ij}(k^*) -1),
\end{equation}
where the $i,j$ denote all possible impurity and feed-down contributions. These
$\lambda$ parameters can be obtained employing exclusively single particle
properties such as the purity and feed-down probability. The underlying
mathematical formalism is outlined in App.~\ref{Seq:AppLambdaParam}.

For the case of \pP correlation the following contributions must be taken into account 
 \begin{equation}
\begin{split}
\{\mathrm{pp}\} = &~ \mathrm{pp+p_{\Lambda}p+ p_{\Lambda}  p_{\Lambda}+p_{\Sigma^+}p+p_{\Sigma^+}p_{\Sigma^+}} \\ 
  & \mathrm{+p_{\Lambda}p_{\Sigma^+}+ \tilde{p}p+ \tilde{p}p_{\Lambda} + \tilde{p}p_{\Sigma^+} + \tilde{p}\tilde{p}},
 \end{split}
 \label{eq:ppCorr}
\end{equation}
where $\tilde{X}$ refers to misidentified particles of specie $X$.
p$_{\Lambda}$ and p$_{\Sigma^+}$ correspond to protons stemming from the weak
decay of the corresponding hyperons. The $\Xi \rightarrow \Lambda \pi
\rightarrow p \pi \pi$ decays are explicitly considered in the feed-down
contribution of the \pL correlation and hence are omitted in Eq.~(\ref{eq:ppCorr}) to avoid double counting. As shown in App.~\ref{Seq:AppLambdaParam}, the fraction of primary protons and their
feed-down fractions  are required to calculate the $\lambda$ parameters of the
different contributions to Eq.~(\ref{eq:ppCorr}). The information about the
origin of the protons, i.e.~whether the particles are of primary origin,
originating from feed-down or from the interactions with the detector material, is
obtained by fitting Monte Carlo (MC) templates to the experimental
distributions of the distance of closest approach of the track to the primary
vertex. The MC templates and the purity are extracted from Pythia
\cite{1126-6708-2006-05-026} simulations using the Perugia 2011 tune
\cite{PhysRevD.82.074018}, which were filtered through the ALICE detector and
the reconstruction algorithm~\cite{1748-0221-3-08-S08002}.
The $p_{\mathrm{T}}$ averages are then calculated by weighting the quantities
of interest by the respective particle yields d$N$/d$p_{\mathrm{T}}$. The resulting fraction of primary protons averaged over
$p_{\mathrm{T}}$ is 87\,\% (where in this fraction we also include the protons
stemming from strong decays of broad resonances), with the other 13\,\% of the total yield associated to weak
decays of resonances and the contribution from the detector material is found to be
negligible. 
 
The feed-down from weakly decaying resonances is evaluated by using cross
sections from Pythia and for the proton sample consists of the $\Lambda$ (70\,\%) and $\Sigma^+$ (30\,\%) contributions. The individual contributions to the total correlation
function are presented in Tab.~\ref{tab:lambdaval}.

The decomposition of the \pL correlation function is conducted in a similar
manner as for the \pP pair, however considering the purities and feed-down
fractions of both particles
\begin{equation}
\begin{split}
\{\mathrm{p}\Lambda\} = &~ \mathrm{p}\Lambda + \mathrm{p}\Lambda_{\Xi^-} + \mathrm{p}\Lambda_{\Xi^0} +
\mathrm{p}\Lambda_{\Sigma^0} + \mathrm{p}_\Lambda \Lambda + \mathrm{p}_\Lambda \Lambda_{\Xi^-} + \mathrm{p}_\Lambda
\Lambda_{\Xi^0} + \mathrm{p}_\Lambda \Lambda_{\Sigma^0} \\
 & +\mathrm{p}_{\Sigma^+} \Lambda + \mathrm{p}_{\Sigma^+} \Lambda_{\Xi^-} +
 \mathrm{p}_{\Sigma^+}\Lambda_{\Xi^0} + \mathrm{p}_{\Sigma^+}\Lambda_{\Sigma^0} +
 \tilde{\mathrm{p}}\Lambda + \tilde{\mathrm{p}}\Lambda_{\Xi^-} + \tilde{\mathrm{p}}\Lambda_{\Xi^0} +  \tilde{\mathrm{p}}\Lambda_{\Sigma^0} \\
& + \mathrm{p}\tilde{\Lambda} +
 \mathrm{p}_\Lambda\tilde{\Lambda}+
 \mathrm{p}_{\Sigma^+}\tilde{\Lambda}+ \tilde{\mathrm{p}}\tilde{\Lambda}.
 \end{split}
\label{eq:ProtonLambdaPairs}
\end{equation}
The \La purity is obtained from fits to the invariant mass spectra in eight
bins of $p_{\mathrm{T}}$ and defined as $S/(S+B)$, where $S$ denotes the actual
signal and $B$ the background. The feed-down contribution is determined from MC
template fits of the experimental distributions of the cosine pointing angle, in
which a total of four templates are considered corresponding to direct,
feed-down, material and impurity contributions. The production probability
d$N$/d$p_{\mathrm{T}}$ is employed in order to obtain $p_{\mathrm{T}}$ weighted
average values. \\
Around $73\%$ of the $\Lambda$s are primaries (where in this fraction we also include the $\Lambda$
stemming from strong decays of broad resonances) and $23\%$ originate from weakly decaying resonances, which is in
line with the values quoted in~\cite{Abbas2013}. The remaining yield is
associated to combinatorial background and $\Lambda$s produced in the detector
material. 
The main contribution to the feed-down fraction is expected to originate from
the $\Xi$ states with no preference for the neutral or the charged,
respectively. This hypothesis is supported by Pythia simulations where the
secondary \La particles arise from the weak decay of the $\Xi^0$ ($48\%$) and
$\Xi^\mathrm{\pm}$ ($49\%$) resonances. 
The remaining contribution in the
simulation arises from the $\Sigma^0$, which however is treated separately.
Since the latter decays electromagnetically almost exclusively into $\Lambda
\gamma$~\cite{Patrignani:2016xqp}, it has a very short life time and cannot be
experimentally differentiated from the sample of primary $\Lambda$s. 
Measurements of the ratio $R_{\Sigma^0/\Lambda}=\sigma_{\Sigma^0}/\sigma_{\Lambda}$ have obtained values around 1/3 \cite{Adamczewski-Musch:2017gjr, 200079, ACCIARRI1994223, 0954-3899-31-6-072}, however with large uncertainties for hadronic collisions at high energies. 
For lack of better estimates the value of 1/3 is used in the following.
The resulting $\lambda$ parameters for the \pL pair
are shown in Tab.~\ref{tab:lambdaval}.

For the \LL correlation function the following pair contributions are taken
into account
\begin{equation}
\begin{split}
 \{ \Lambda \Lambda \} =&~\Lambda \Lambda + \Lambda \Lambda_{\Sigma^0} +
 \Lambda_{\Sigma^0} \Lambda_{\Sigma^0} + \Lambda \Lambda_{\Xi^0} +
 \Lambda_{\Xi^0} \Lambda_{\Xi^0} + \Lambda \Lambda_{\Xi^-} \\
 & + \Lambda_{\Xi^-} \Lambda_{\Xi^-} + \Lambda_{\Sigma^0} \Lambda_{\Xi^0} +
 \Lambda_{\Sigma^0} \Lambda_{\Xi^-} + \Lambda_{\Xi^0} \Lambda_{\Xi^-} \\
& + \tilde{\Lambda} \Lambda +
 \tilde{\Lambda} \Lambda_{\Sigma^0} +
  \tilde{\Lambda} \Lambda_{\Xi^-} +
   \tilde{\Lambda} \Lambda_{\Xi^0} + 
  \tilde{\Lambda} \tilde{\Lambda}.
\end{split}
\end{equation}
The resulting $\lambda$ parameters are shown in Tab.~\ref{tab:lambdaval}.
Notable is that the actual pair of interest contributes only to about one third
of the signal, while pair fractions involving in particular $\Sigma^0$ and
$\Xi$ give a significant contribution. The statistical uncertainties of these
parameters are negligible and their influence on the systematic uncertainties will be evaluated
in Sec.~\ref{Sec:Systematics}. 
\begin{table}
\begin{center}
\begin{tabularx}{\textwidth}{l|X || l|X || l|X}
\hline  \hline
\multicolumn{2}{c||}{\pP} & 
\multicolumn{2}{c||}{\pL} & 
\multicolumn{2}{c}{\LL} \\

Pair & $\lambda$ parameter [\%] &
Pair & $\lambda$ parameter [\%] &
Pair & $\lambda$ parameter [\%] \\
\hline 
pp & 74.18 &
p$ \Lambda$ & 47.13 &
$\Lambda \Lambda$ & 29.94\\

p$_{\Lambda}$p & 15.52 &
p$ \Lambda_{\Xi^{-}}$ & 9.92&
$\Lambda \Lambda_{\Sigma^{0}}$ & 19.96\\

p$_{\Lambda}$p$_{\Lambda}$ & 0.81&
p$ \Lambda_{\Xi^{0}}$ & 9.92&
$\Lambda_{\Sigma^{0}} \Lambda_{\Sigma^{0}}$ & 3.33\\

p$_{\Sigma^+}$p & 6.65 &
p$ \Lambda_{\Sigma^{0}}$ & 15.71 &
$\Lambda \Lambda_{\Xi^{0}}$ & 12.61\\

p$_{\Sigma^+}$p$_{\Sigma^+}$ & 0.15 &
p$_{\Lambda} \Lambda$ & 4.93 &
$\Lambda_{\Xi^{0}} \Lambda_{\Xi^{0}}$ & 1.33\\

p$_{\Lambda}$p$_{\Sigma^+}$ & 0.70&
p$_{\Lambda} \Lambda_{\Xi^-}$ & 1.04&
$\Lambda \Lambda_{\Xi^{-}}$ & 12.61\\

$\tilde{\mathrm{p}}$p & 1.72 &
p$_{\Lambda} \Lambda_{\Xi^0}$ & 1.04 &
$\Lambda_{\Xi^{-}} \Lambda_{\Xi^{-}}$ & 1.33\\

$\tilde{\mathrm{p}}$p$_{\Lambda}$  & 0.18 &
p$_{\Lambda} \Lambda_{\Sigma^0}$ & 1.64 &
$\Lambda_{\Sigma^{0}} \Lambda_{\Xi^{0}}$ & 4.20\\

$\tilde{\mathrm{p}}$p$_{\Sigma^+}$  & 0.08 &
p$_{\Sigma^+} \Lambda$ &  2.11 &
$\Lambda_{\Sigma^{0}} \Lambda_{\Xi^{-}}$ & 4.20\\

$\tilde{\mathrm{p}}\tilde{\mathrm{p}}$ & 0.01 &
p$_{\Sigma^+} \Lambda_{\Xi^-}$ & 0.44 &
$\Lambda_{\Xi^{0}} \Lambda_{\Xi^{-}}$ & 2.65\\

&&
p$_{\Sigma^+} \Lambda_{\Xi^0}$ & 0.44 &
$\tilde{\Lambda} \Lambda$ & 4.38\\

&&
p$_{\Sigma^+} \Lambda_{\Sigma^0}$ & 0.70 &
$\tilde{\Lambda} \Lambda_{\Sigma^{0}}$ & 1.46 \\

&&
$\tilde{\mathrm{p}} \Lambda$ & 0.55 &
$\tilde{\Lambda} \Lambda_{\Xi^{0}}$ & 0.92 \\

&&
$\tilde{\mathrm{p}} \Lambda_{\Xi^-}$ & 0.18 &
$\tilde{\Lambda} \Lambda_{\Xi^{-}}$ & 0.92 \\

&&
$\tilde{\mathrm{p}} \Lambda_{\Xi^0}$ & 0.12 &
$\tilde{\Lambda} \tilde{\Lambda}$ & 0.16\\

&&
$\tilde{\mathrm{p}} \Lambda_{\Sigma^0}$ & 0.12 &
& \\

&&
p$ \tilde{\Lambda}$ & 3.45 &
& \\
&&
p$_{\Lambda} \tilde{\Lambda}$ & 0.36 &
& \\

&&
p$_{\Sigma^+} \tilde{\Lambda}$ & 0.15 &
& \\

&&
$\tilde{\mathrm{p}} \tilde{\Lambda}$ & 0.04 &
& \\
\hline \hline
\end{tabularx}
\caption[lambda parameters]{Weight parameters of the individual components of the \pP, \pL and \LL correlation function.}
\label{tab:lambdaval}
\end{center}
\end{table}

Possible effects that were considered in this analysis and that could be influencing the source are either the decay of strong resonances, or a $m_{\mathrm{T}}$ scaling. The latter is related to collective effects and may result in a non-Gaussian behavior of the source. The \pP correlation function is here considered as a benchmark since the theoretical description of the interaction is well established. The good agreement between the $m_{\mathrm{T}}$-integrated data and the femtoscopic fit demonstrates that the assumption of a Gaussian source is pertinent. A comparison of the $m_{\mathrm{T}}$ distribution for \pP and \pL pairs shows a good agreement between the two. The limited experimental sample, however, does not allow conducting a differential analysis yet.\\
Additionally, we have considered the effect of the strong decays such as $\Delta\rightarrow N+\pi$  and $N^* \rightarrow \Lambda + K$ on the production of protons and $\Lambda$. Since the $\Delta$ and $N^*$ resonances have typical widths above 120\,MeV the decay length is in the order of 1\,fm so that the decay particles do still experience the final state interaction with the neighbouring particles as the primaries do. We estimate that about 65\% of all primary protons and $\Lambda$ stem from the strong decay of resonances \cite{Becattini:2009ee} and  we have simulated
 how the source could be modified by such effects. A difference of 5\% in the results of the Gaussian fit is found when comparing \pP to \pL pairs. These effects are in the same order of the present uncertainty on the radius and are herewith neglected. A quantitative study including all resonances is planned in the analysis of the LHC Run 2 sample.

\subsection{Detector effects}

The shape of the experimentally determined correlation function is affected by
the finite momentum resolution.
This is taken into account when the experimental data are compared to model
calculations in the fitting procedure by transforming the modeled correlation
function, see Eq.~(\ref{eq:totcorrelation}), to the reconstructed momentum basis.

When tracks of particle pairs involved in the correlation function are almost 
collinear, i.e. have a low $k^*$, detector effects can affect the measurement.
No hint for track merging or splitting is found and therefore no explicit
selection criteria are introduced.

\subsection{Non-femtoscopic background}   
\label{Seq:NonFemtoBackground}
 
\begin{figure*}
\centering
\includegraphics[width=\textwidth]{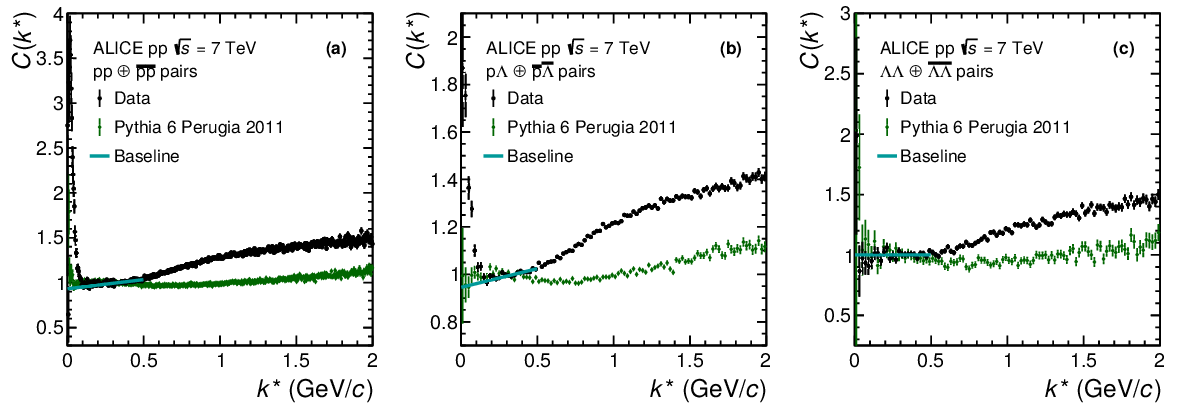}
\caption{\textit{(Color online)} The raw correlation function compared to
Pythia 6 Perugia 2011 simulations for \pP (a), \pL (b) and \LL (c) pairs.}
\label{fig:CFPythia}
\end{figure*}

For sufficiently large relative momenta ($k^*>200$\,MeV/$c$) and increasing separation distance, the FSI among the
particles is suppressed and hence the correlation function should approach unity.
As shown in Fig.~\ref{fig:CFPythia}, however, the measured correlation function
for \pP and \pL exhibits an increase for $k^*$ larger than about
$200$\,MeV/$c$ for the two systems. Such
non-femtoscopic effects, probably due to energy-momentum conservation, are in
general more pronounced in small colliding systems where the average particle
multiplicity is low ~\cite{Lisa:2005dd}. In the case of meson\mbox{--}meson
correlations at ultra-relativistic energies, the appearance of long-range
structures in the correlation functions for moderately small $k^*$ ($k^* < 200$\,MeV/\textit{c}) is typically
interpreted as originating from mini-jet-like structures
~\cite{PhysRevD.84.112004, PhysRevC.91.034906}.

Pythia also shows the same non-femtoscopic correlation for larger $k^*$ but
fails to reproduce quantitatively the behavior shown in
Fig.~\ref{fig:CFPythia}, as already observed for the angular correlation of
baryon\mbox{--}baryon and anti-baryon\mbox{--}anti-baryon pairs~\cite{Adam2017}. 

Energy-momentum conservation leads to a contribution to the signal which can be
reproduced with a formalism described in~\cite{Bock:2011} and is accordingly also
considered in this work. Therefore, a linear function
$C(k^*)_{\mathrm{non-femto}}=a k^* + b$ where $a,b$ are fit parameters, is
included to the global fit as $ C(k^*)=C(k^*)_{\mathrm{femto}} \times
C(k^*)_{\mathrm{non-femto}}$ to improve the description of the signal by the
femtoscopic model. The fit parameters of the baseline function are obtained in
$k^* \in [0.3, ~0.5]$\,GeV/$c$ for \pP and \pL pairs. For the case of the \LL correlation function, the uncertainties of the data do not allow to additionally add a baseline, which is therefore omitted in the femtoscopic fit.

\subsection{Modeling the correlation function}
\label{deco:models}

\subsubsection{Genuine correlation function}

For the \pP correlation function the Coulomb and the strong interaction as well
as the antisymmetrization of the wave functions are considered
\cite{KOONIN197743}. The strong interaction part of the potential is modeled
employing the Argonne $v_{18}$~\cite{PhysRevC.51.38} potential considering the $s$ and $p$ waves. The source is assumed to be isotropic with a Gaussian profile of radius $r_0$. The resulting Schr\"{o}dinger equation is then solved with the CATS~\cite{CATS}.
 
In the case of \pL and \LL we employ the Lednick\'{y} and Lyuboshitz
analytical model~\cite{Ledn} to describe these correlation functions. This
model is based on the assumption of an isotropic source with Gaussian profile
\begin{equation}
S(r_0) = \frac{1}{(4\pi r_0 ^2)^{3/2}} \text{exp}  \left( -\frac{r^2}{4r_0 ^2 } \right),
\end{equation}
where $r_0$ is the size of the source.
Additionally, the complex scattering amplitude is evaluated by means of the effective
range approximation
\begin{equation}
f(k^*)^{S}=\left( \frac{1}{f_0^S}+\frac{1}{2}d_0^S k^{*2} -ik^* \right)^{-1} , 
\end{equation}
with the scattering length $f_0^S$, the effective range $d_0^S$ and $S$
denoting the total spin of the particle pair. In the following, the usual sign
convention of femtoscopy is employed where an attractive interaction leads to
a positive scattering length. With these assumptions the analytical description
of the correlation function for uncharged particles~\cite{Ledn} reads 
\begin{equation}
C(k^*)_{\rm{Lednicky}}=1+\sum_S \rho_S \left[ \frac{1}{2} \left|
\frac{f(k^*)^S}{r_0} \right|^2 \left( 1-\frac{d_0^S}{2\sqrt{\pi} r_0} \right) +
\frac{2 \Re{f(k^*)^S}}{\sqrt{\pi} r_0} F_1(Q_{\rm{inv}}r_0) -
\frac{\Im{f(k^*)^S}}{r_0} F_2(Q_{\rm{inv}}r_0) \right],
\label{eq:CFgauss}
\end{equation}
where $\Re{f(k^*)^S}$ ($\Im{f(k^*)^S}$) denotes the real (imaginary) part of
the complex scattering amplitude, respectively. The $F_1(Q_{\rm{inv}}r_0)$ and
$F_2(Q_{\rm{inv}}r_0)$ are analytical functions resulting from the
approximation of isotropic emission with a Gaussian source and the factor
$\rho_S$ contains the pair fraction emitted into a certain spin state $S$. For
the \pL pair unpolarized emission is assumed.

The \LL pair is composed of identical particles and hence additionally quantum
statistics needs to be considered, which leads to the introduction of an
additional term to the Lednick\'{y} model, as employed e.g.~in
\cite{Adamczyk:2014vca}.

While the CATS framework can provide an exact solution for any source and local interaction potential, the Lednicky-Lyuboshitz approach uses the known analytical solution outside the range of the strong interaction potential and takes into account its modification in the inner region in an approximate way only. That is why this approach may not be valid for small systems. 
\subsubsection{Residual correlations} 
\label{Sec:Residuals}

Table~\ref{tab:lambdaval} demonstrates that a significant admixture of residuals is present in the experimental sample of particle pairs. 
A first theoretical investigation of these so-called residual correlations was conducted in \cite{PhysRevC.60.067901}.
This analysis relies on the procedure established in \cite{Kisiel:2014mma}, where the initial correlation function of the residual is calculated and then transformed to the new momentum basis after the decay.



For the \pP channel only the feed-down from the \pL correlation function is
considered, which is obtained by fitting the \pL experimental correlation
function and then transforming it to the \pP momentum basis. All contributions
are weighted by the corresponding $\lambda$ parameters and the modeled
correlation function for this pair $C_{model, \mathrm{p\mbox{--}p}}(k^*)$ can be written as
\begin{equation}
C_{model, \mathrm{p\mbox{--}p}}(k^*) = 1 + \lambda_{\mathrm{pp}}  \cdot ( C_{\mathrm{pp}}(k^*)-1 ) +\lambda_{\mathrm{pp}_{\Lambda}} (C_{\mathrm{pp}_{\Lambda}}(k^*) -1).
\end{equation}
All other residual correlations are assumed to be flat.

For the \pL, residual correlations from the \pSiZero, \pXim and \LL pairs are
taken into account. As the \LL correlation function is rather flat no further
transformation is applied. The \pSiZero correlation function is obtained using
predictions from~\cite{Stavinskiy:2007wb}. 

As the decay products of the reaction $\Xi \rightarrow \Lambda \pi$ are charged
and therefore accessible by ALICE, we measure the p\mbox{--}$\Xi$ correlation
function. The experimental data are parametrized with a phenomenological
function
\begin{equation}
C_{\mathrm{p\mbox{--}}\Xi^-}(k^*)=1+\frac{\exp(-k^* a_\Xi)}{k^* a_\Xi},
\end{equation}
where the parameter $a_\Xi$ is employed to scale the function to the data and has no
physical meaning. Its value is found to be $a_\Xi=3.88$\,fm.

The modeled correlation function $C_{model,\mathrm{p}\mbox{--}\Lambda}(k^*)$ for the pair is
obtained by
\begin{equation}
 C_{model, \mathrm{p}\mbox{--}\Lambda}(k^*)= 1+\lambda_{\mathrm{p}\Lambda} (C_{\mathrm{p}\Lambda}(k^*)-1) +
 \lambda_{\mathrm{p}\Lambda_{\Sigma^0}} (C_{\mathrm{p}\Lambda_{\Sigma^0}}(k^*)-1) +
 \lambda_{\mathrm{p}\Lambda_{\Xi^-}} (C_{\mathrm{p}\Lambda_{\Xi^-}}(k^*)-1). 
\end{equation}

As the present knowledge on the hyperon\mbox{--}hyperon interaction is scarce, in
particular regarding the interaction of the $\Lambda$ with other hyperons, all
residual correlations feeding into the \LL correlation function are considered
to be consistent with unity,
\begin{equation}
 C_{model, \Lambda\mbox{--}\Lambda}(k^*) = 1+\lambda_{\Lambda\Lambda} (C_{\Lambda\Lambda}(k^*)-1).
\end{equation}

It should be noted, that the residual correlation functions, after weighting
with the corresponding $\lambda$ parameter, transformation to the momentum base
of the correlation of interest and taking into account the finite momentum
resolution, only barely contribute to the total fit function.

\subsubsection{Total correlation function model}
\label{Seq:TotCorr}
The correlation function modeled according to the considerations discussed
above is then multiplied by a linear function to correct for the baseline as
discussed in Sec.~\ref{Seq:NonFemtoBackground} and weighted with a normalization
parameter $\mathcal{N}$ 
\begin{equation}
C_{tot}(k^*) = \mathcal{N}\cdot (a+b\cdot k^*)\cdot C_{model}(k^*),
\label{eq:totcorrelation}
\end{equation}
where $C_{model}(k^*)$ incorporates all considered theoretical correlation
functions, weighted with the corresponding $\lambda$ parameters as discussed in
Sec.~\ref{deco} and \ref{deco:models}. 

The inclusion of a baseline is further motivated by the presence of a linear but non-flat correlation 
observed in the data outside the femtoscopic region (see Fig.~\ref{fig:CFPythia} for $k^*\in[0.3,~0.5]$\,GeV/$c$). 
When attempting to use a higher order polynomial to model the background, the resulting curves are still 
compatible with a linear function, while their interpolation into the lower $k^*$ region leads to an overall 
poorer fit quality.
\section{Systematic uncertainties}
\label{Sec:Systematics}
\subsection{Correlation function}
The systematic uncertainties of the correlation functions are extracted by
varying the proton and \La candi-date selection criteria according to
Tab.~\ref{tab:SysErrorbudget}. Due to the low number of particle pairs, in
particular at low $k^*$, the resulting variations of the correlation functions
are in general much smaller than the statistical uncertainties. In order to still estimate the systematic
uncertainties the data are rebinned by a factor of 10. The systematic
uncertainty on the correlation function is obtained by computing the ratio of
the default correlation function to the one obtained by the respective cut
variation. Whenever this results in two systematic uncertainties, i.e.~by a
variation up and downwards, the average is taken into account. Then all
systematic uncertainties from the cut variations are summed up quadratically.
This is then extrapolated to the finer binning of the correlation function by fitting a polynomial of
second order. The obtained systematic uncertainties are found to be largest in the lowest $k^*$ bin.
The individual contributions in that bin are
summarized in Tab.~\ref{tab:SysErrorbudget} and the resulting total systematic
uncertainty accounts to about 4\,\% for \pP, 1\,\% for \pL and 2.5\,\% for \LL.
Variations of the proton DCA selection are not taken into account for the
computation of the systematic uncertainty since it dilutes (enhances) the correlation
signal by introducing more (less) secondaries in the sample. This effect is recaptured by a change
in the $\lambda$ parameter.

     
\begin{table}
\begin{center}     
\begin{tabularx}{1.\textwidth}{b|b|b|c|c|c}
\hline  \hline
\heading{Variable} & \heading{Default} & \heading{Variation} & \heading{\pP
[\%]} & \heading{\pL [\%]} & \heading{\LL [\%]} \\
\hline 
\cline{1-6} Min. $p_{\mathrm{T}}$ proton (\GeVc) & 	0.5 & 0.4, 0.6 & 1 & 0.2 &
- \\
\cline{1-6} $|\eta|$ proton & 0.8 & 0.7, 0.9	& 0.4 	& 0.2	& -\\
\cline{1-6} $n_{\sigma}$ proton & 3 & 2, 5	& 1.8		& 0.2	& -\\
\cline{1-6} Proton tracks & TPC only & Global 	& 2.4	& 0		& -\\
\cline{1-6} $n_{\mathrm{Cluster}}$ proton 	& 80 & 90 & 0.3	& 0.1		& -\\
\cline{1-6} Min. $p_{\mathrm{T}}$ $V_0$ (\GeVc) & 0.3 & 0.24, 0.36 & - & 0 & 0 \\
\cline{1-6} cos$(\alpha)$ $V_0$ & 0.99 & 0.998 & - & 0 & 1.8 \\
\cline{1-6} $n_{\sigma}$ $V_0$ daughter	& 5 & 4 & - & 0.1 & 0.3 \\
\cline{1-6} $n_{\mathrm{Cluster}}$ $V_0$ daughter & 70 & 80 & - & 0.1		& 0.7 \\
\cline{1-6} $|\eta|$ $V_0$ & 0.8 & 0.7, 0.9	& - & 0.6 & 0.8 \\
\cline{1-6} $\mathrm{DCA}(|p,\pi|)$ (cm) & 1.5 & 1.2 & - & 0.5	& 0 \\
\cline{1-6} $\mathrm{DCA}$ (cm) & 0.05 & 0.06 & - & 0.7 & 0.6 \\
\hline \hline
\end{tabularx}
\caption[Systematic uncertainties]{Selection parameter variation and the resulting relative systematic uncertainty on the \pP, \pL and \LL correlation function.}
\label{tab:SysErrorbudget}
\end{center}
\end{table}

\subsection{Femtoscopic fit}
To evaluate the systematic uncertainty of the femtoscopic fit, and hence on the
measurement of the radius $r_0$, the fit is performed applying the following variations. 
Instead of the common fit, the radius is determined separately from
the \pP and \pL correlation functions. \LL is excluded because it imposes only a shallow constraint on the radius, in particular since the scattering parameters are unconstrained for the fit.
 Furthermore, the input to
the $\lambda$ parameters are varied by 25\,\%, while keeping the purity and the
fraction of primaries and secondaries constant since this would correspond to a
variation of the particle selection and thus would require a different
experimental sample as discussed above. Additionally, all fit ranges of both
the femtoscopic and the baseline fits are varied individually by up to 50\,\%
and 10\,\%, respectively. The lower bound of the femtoscopic fit is always left
at its default value. For the \pL correlation function the dependence on the
fit model is studied by replacing the Lednick\'{y} and Lyuboshitz analytical
model with the potential introduced by Bodmer, Usmani, and Carlson
\cite{PhysRevC.29.684} for which the Schr\"{o}dinger equation  is explicitly
solved using CATS. Additionally, the fit for the \pP and \pL correlation
function is performed without the linear baseline. The radius is determined for
2000 random combinations of the above mentioned variations. The resulting
distribution of radii is not symmetric and the systematic uncertainty is
therefore extracted as the boundaries of the 68\,\% confidence interval around
the median of the distribution and accounts to about 4\,\% of the determined
radius.

\section{Results}
\label{Sec:Results}

\begin{figure*}
\centering
\includegraphics[width=1\textwidth]{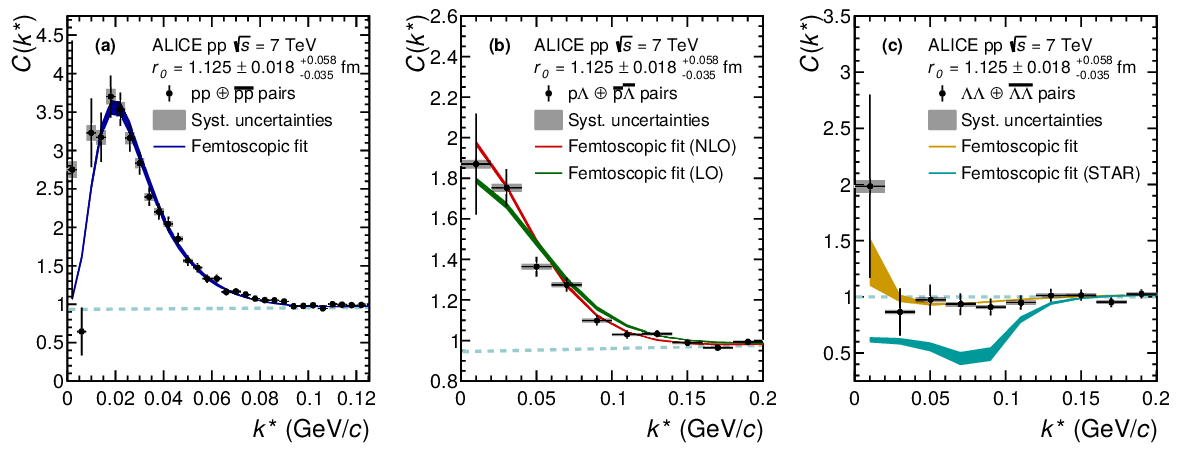}
\caption{\textit{(Color online)} The \pP (a), \pL (b) and \LL (c) correlation function with a simultaneous fit with the NLO expansion (red line) for the scattering parameter of \pL~\cite{Haidenbauer:2013oca}. The dashed line denotes the linear baseline. After the fit is performed the LO~\cite{POLINDER2006244} parameter set (green curve) is plugged in for the \pL system and the scattering length obtained from~\cite{Adamczyk:2014vca} for the \LL system (cyan curve).}
\label{fig:PPLPfit}
\end{figure*}

The obtained \pP, \pL and \LL correlation functions are shown in Fig.~\ref{fig:PPLPfit}. 
For each of the correlation functions we do not observe any mini-jet background in the low $k^*$ region, as observed in the case of neutral~\cite{2012151} and charged~\cite{Abelev:2012sq} kaon pairs and charged pion pairs~\cite{PhysRevD.84.112004}.
This demonstrates that the femtoscopic signal in baryon\mbox{--}baryon correlations is dominant in ultrarelativistic \pp collisions.
The signal amplitude for the \pP and \pL correlations are much larger than the one observed in analogous studies from heavy-ion collisions~\cite{Pratt:1986cc, Adams:2005ws, Anticic:2011ja, Agakishiev:2010qe}, due to the small particle emitting source formed in \pp collisions, allowing a higher sensitivity to the FSI.

In absence of residual contributions and any FSI, the \LL correlation function is expected to approach 0.5 as $k^*$ $\rightarrow$ 0.
While the data suggest that the \LL correlation exceeds the value expected considering only quantum statistic effects, the limited amount of data of the herewith presented sample does not allow to draw strong conclusions on the attractive nature of the \LL interaction~\cite{Hashimoto:2006aw, PhysRevC.91.024916}.

The experimental data are fitted using CATS and hence the exact solution of the Schr\"odinger equation for the \pp correlation and the 
Lednick\'{y} model for the \pL and \LL correlation. The three fits are done simultaneously and this way the source radius is extracted 
and different scattering parameters for the \pL and \LL interactions can be tested.
While in the case of the \pP and \pL correlation function the existence of a baseline is clearly visible in the data, the low amount of pairs in the \LL channel do not allow
for such a conclusion. Therefore, the baseline is not included in the model for the \LL correlation function.

The simultaneous fit is carried out by using a combined $\chi^2$ and with the radius as a free parameter common to all correlation
functions. The fit range is $k^*$ $\in [0,~0.16]$\,GeV/$c$ for \pP and $k^*$ $\in [0,~0.22]$\,GeV/$c$ for \pL and \LL. 
Hereafter we adopt the convention of positive scattering lengths for attractive interactions and negative scattering lengths for repulsive interactions.
The \pL strong interaction is modeled employing scattering parameters obtained using the next-to-leading order expansion of a chiral effective field theory at
a cutoff scale of $\Lambda=600$\,MeV~\cite{Haidenbauer:2013oca}.  
The simultaneous fit of the \pP, \pL and \LL correlation functions yields a common radius of \radiusResult. 

The blue line in the left panel in Fig.~\ref{fig:PPLPfit} shows the result of the femtoscopic fit to the \pP correlation function using the Argonne $v_{18}$ potential that describes the experimental data in a satisfactory way.
The red curve in the central panel shows the result of the NLO calculation for \pL.
In the case of \LL (right panel), the yellow curve represents the femtoscopic fit with free scattering parameters.
The width of the femtoscopic fits corresponds to the systematic uncertainty of the correlation function discussed in Sec.~\ref{Sec:Systematics}.

After the fit with the NLO scattering parameters has converged, the \pL correlation function for the same source size is compared to the data using various theoretically obtained scattering parameters~\cite{ND,NF,NSC89,NSC97,ESC08,POLINDER2006244,Haidenbauer:2013oca, JulichA, JulichJ} as summarized in Tab.~\ref{tab:pLscattering}. The degree of consistency is expressed in the number of standard deviations $n_{\sigma}$.
The employed models include several versions of meson exchange models proposed such as the Nijmegen model D (ND)~\cite{ND}, model F (NF)~\cite{NF}, soft core (NSC89 and NSC97)~\cite{NSC89, NSC97} and extended soft core (ESC08)~\cite{ESC08}.
Additionally, models considering contributions from one- and two-pseudoscalar-meson exchange diagrams and from four-baryon contact terms in $\chi$EFT at leading~\cite{POLINDER2006244} and next-to-leading order~\cite{Haidenbauer:2013oca} are employed, together with the first version of the J\"ulich Y\mbox{--}N meson exchange model~\cite{JulichA}, which in a later version~\cite{JulichJ} also features one-boson exchange.
 
All employed models describe the data equally well and hence the available data does not allow yet for a discrimination.
As an example, we show in the central panel of Fig.~\ref{fig:PPLPfit} how employing scattering parameters different than the NLO ones reflects on the \pL correlation function.
The green curve corresponds to the results obtained employing LO scattering parameters and the theoretical correlation function is clearly sensitive for $k^*$ $\rightarrow$ 0 to the input parameter. 

\begin{table}
\begin{center} 
\begin{tabularx}{1.\textwidth}{XX|X|X|X|X|X} 
\hline  \hline
Model & &$f_0^{S=0}$\,(fm) & $f_0^{S=1}$\,(fm) & $d_0^{S=0}$\,(fm) & $d_0^{S=1}$\,(fm)  & $n_\sigma$ \\
\hline 
\cline{1-7} ND~\cite{ND} && 1.77 & 2.06 & 3.78 & 3.18&  1.1\\

\cline{1-7} NF~\cite{NF} && 2.18 & 1.93 & 3.19 & 3.358 & 1.1\\

\cline{1-7} NSC89~\cite{NSC89} && 2.73 & 1.48 & 2.87 & 3.04 & 0.9 \\

\cline{1-7} \multirow{6}{*}{NSC97~\cite{NSC97}} &a& 0.71 & 2.18 & 5.86 & 2.76 & 1.0 \\
\cline{2-7} & b & 0.9 & 2.13 & 4.92 & 2.84 & 1.0 \\
\cline{2-7} & c & 1.2 & 2.08 & 4.11 & 2.92 & 1.0\\
\cline{2-7} & d & 1.71 & 1.95 & 3.46 & 3.08 & 1.0\\
\cline{2-7} & e & 2.1 & 1.86 & 3.19 & 3.19 & 1.1 \\
\cline{2-7} & f &  2.51 & 1.75 & 3.03 & 3.32 &  1.0\\

\cline{1-7} ESC08~\cite{ESC08} && 2.7 & 1.65 &2.97 & 3.63 & 0.9 \\

\cline{1-7} \multirow{2}{*}{$\chi$EFT} & LO~\cite{POLINDER2006244}  & 1.91 & 1.23 & 1.4 & 2.13 & 1.8\\

\cline{2-7}  &  NLO~\cite{Haidenbauer:2013oca} & 2.91 & 1.54 & 2.78 & 2.72 & 1.5 \\

\cline{1-7}  \multirow{3}{*}{J\"ulich} & A~\cite{JulichA} & 1.56 & 1.59 & 1.43  & 3.16 & 1.0 \\
\cline{2-7} & J04~\cite{JulichJ} & 2.56 & 1.66 & 2.75 & 2.93 & 1.4 \\
\cline{2-7} & J04c~\cite{JulichJ} & 2.66  & 1.57 &  2.67 & 3.08 & 1.1 \\
\hline \hline
\end{tabularx}
\caption{Scattering parameters for the \pL system from various theoretical calculations~\cite{ND,NF, NSC89,NSC97,ESC08,POLINDER2006244,Haidenbauer:2013oca, JulichA, JulichJ} and the corresponding degree of consistency with the experimentally determined correlation function expressed in numbers of standard deviations $n_\sigma$. The $\chi$EFT scattering parameters are obtained at a cutoff scale $\Lambda = 600$\,MeV. The usual sign convention of femtoscopy is employed where an attractive interaction leads to a positive scattering length.}
\label{tab:pLscattering}
\end{center}
\end{table}

\begin{figure}[t]
\centering
\includegraphics[width=0.9\textwidth]{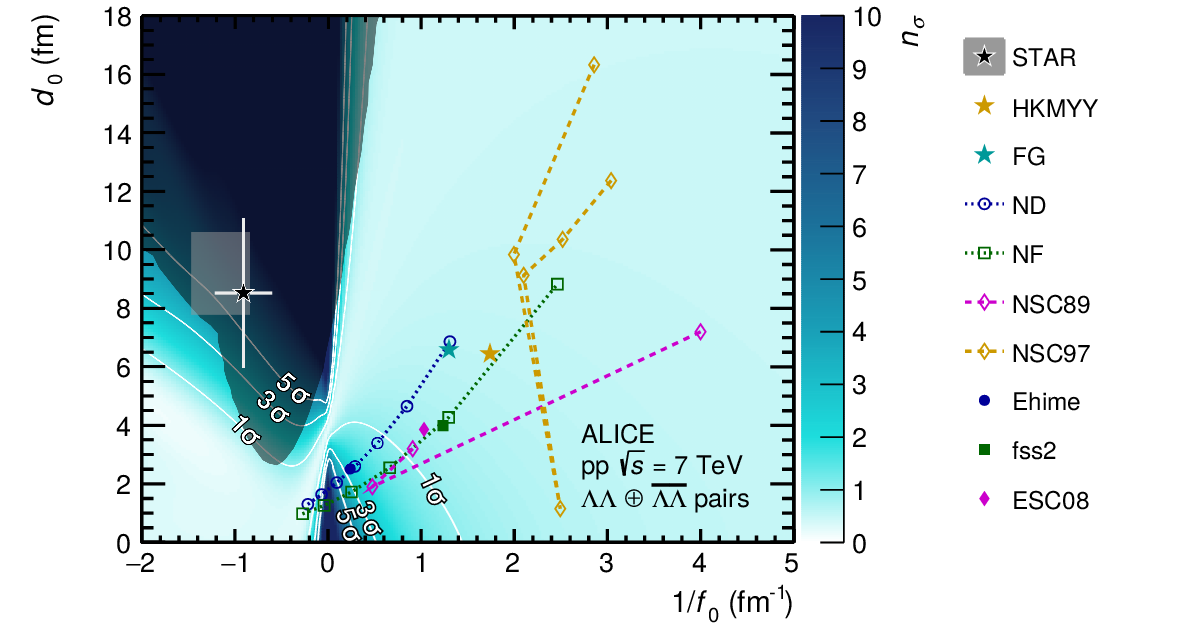}
\caption{\textit{(Color online)} Number of standard deviations $n_{\sigma}$  of the modeled correlation function for a given set of scattering parameters (effective range $d_0$
and scattering length $f_0$)
with respect to the data, together with various model calculations~\cite{ND, NF, NSC89, NSC97, ESC08, Ehime1, Ehime2, fss1, fss2, FG, HKMYY} and measurements~\cite{Adamczyk:2014vca}. The gray shaded area corresponds to the region where the Lednick\'{y} model predicts a negative correlation function for \pp collisions at $\sqrt{s} = 7\,$TeV.}
\label{fig:LambdaLambda}
\end{figure}

In order to probe which scattering parameters are compatible with the measured 
\LL correlation function, the effective range and the scattering length of the
potential are varied within $d_0 \in [0, 18]$\,fm and $1/f_0 \in
[-2, 5]$\,1/fm, while keeping the renormalization constant $\mathcal{N}$ as the
only free fit parameter. It should be noted that the resulting variations of $\mathcal{N}$ are on the percent level.
The resulting correlation functions obtained by employing the Lednick\'{y} and Lyuboshitz
analytical model~\cite{Ledn} and considering also the secondaries and impurities contributions are compared
to the data. The degree of consistency is expressed in the number of standard
deviations $n_{\sigma}$, as displayed in Fig.~\ref{fig:LambdaLambda} together with an overview of the present knowledge
about the \La\mbox{--}\La interaction. 
For a detailed overview of the currently available models see e.g.
\cite{PhysRevC.91.024916}, from which we have obtained the collection of
scattering parameters.
Additionally to the Nijmegen meson exchange models mentioned above, the data
are compared to various other theoretical calculations.
An exemplary boson-exchange potential is Ehime~\cite{Ehime1,Ehime2}, whose
strength is fitted to the outdated double hypernuclear bound energy, $\Delta
B_{\Lambda\Lambda}  = 4$\,MeV~\cite{DANYSZ1963121} and accordingly known to be
too attractive. As an exemplary quark model including baryon\mbox{--}baryon
interactions with meson exchange effects, the fss2 model~\cite{fss1,fss2} is
used.
Moreover, the potentials by Filikhin and Gal (FG)~\cite{FG} and by Hiyama,
Kamimura, Motoba, Yamada, and Yamamoto (HKMYY)~\cite{HKMYY}, which are capable
of describing the NAGARA event~\cite{Takahashi:2001nm} are employed.

In contrast to the \pL case, the agreement with the data increases with every revision of the Nijmegen potential, while the
introduction of the extended soft core slightly increases the deviation. In
particular solution NSC97f yields the overall best agreement with the data. 
The correlation function modeled using scattering parameters of the Ehime model which is known to be too attractive deviates
by about 2 standard deviations from the data. \\
 For an attractive interaction (positive $f_0$) the correlation function is pushed from the quantum statistics distribution for two fermions 
(correlation function equal to $0.5$ for $k^* = \,0$) to unity. As a result within the current uncertainties the \LL correlation function is rather flat and close to 1 and this lack of structure makes it 
impossible to extract the two scattering parameters with a reasonable uncertainty.
This means that even by increasing the data by a factor $10$, as expected from the RUN2 data, it will be very complicated to constrain precisely the region $f_0>0$.\\
As for the region of negative scattering length $f_0$ this is connected in scattering theory either to a repulsive interaction or to the 
existence of a bound state close 
to the threshold and a change in the sign of the scattering length. Since the \LL interaction is known to be slightly attractive above the threshold 
\cite{Takahashi:2001nm}, the measurement of a negative 
scattering lengths would strongly support the existence of the H-dibaryon.
Notably the correlation function modeled employing the scattering parameters obtained by the STAR collaboration in Au\mbox{--}Au collisions at $
\sqrt{s_{\mathrm{NN}}} = 200$\,GeV~\cite{Adamczyk:2014vca} and all the secondaries and impurities contributions deviates by 6.8 standard deviations from the data.
This is also shown by the cyan curve displayed in the right panel of Fig.~\ref{fig:PPLPfit} which is obtained using the source radius and the $\lambda$ 
parameters from this analysis and the scattering parameters from~\cite{Adamczyk:2014vca}.
On the other hand these parameters and all those corresponding to the gray-shaded area in Fig.~\ref{fig:LambdaLambda} lead to a negative genuine \LL correlation function if the Lednick\'{y} model is employed. The total correlation function that is compared to the experimental data is not 
negative because the impurities and secondaries contributions lead to a total correlation function that is always positive.
This means that the combination of large effective ranges and negative scattering lengths translate into unphysical correlation functions, for small colliding 
systems as $\mathrm {p\kern-0.05em p}$. This effect is not immediate visible in larger colliding system such as Au\mbox{--}Au at $\sqrt{s_{\mathrm{NN}}} = 200$\,GeV measured by STAR, where the 
obtained correlation function does not become negative.
This demonstrates that these scattering parameters intervals combined with the Lednick\'{y} model are not suited
to describe the correlations functions measured in small systems.
One could test the corresponding local potentials with the help of CATS \cite{CATS}, since the latter does not suffer from the limitations of the Lednick\'{y} model due to the 
employment of the asymptotic solution. On the other hand we have directly compared the correlation functions obtained employing CATS and the \LL 
local potentials reported in \cite{PhysRevC.91.024916} with the correlation functions obtained using the corresponding scattering parameters and the Lednick\'{y} model.
For the typical source radii of $1.3$\,fm the deviations are within 10\,\%. This disfavours the region of negative scattering lengths and large effective ranges
for the \LL correlation.

\begin{figure}[t]
\centering
\includegraphics[width=0.7\textwidth]{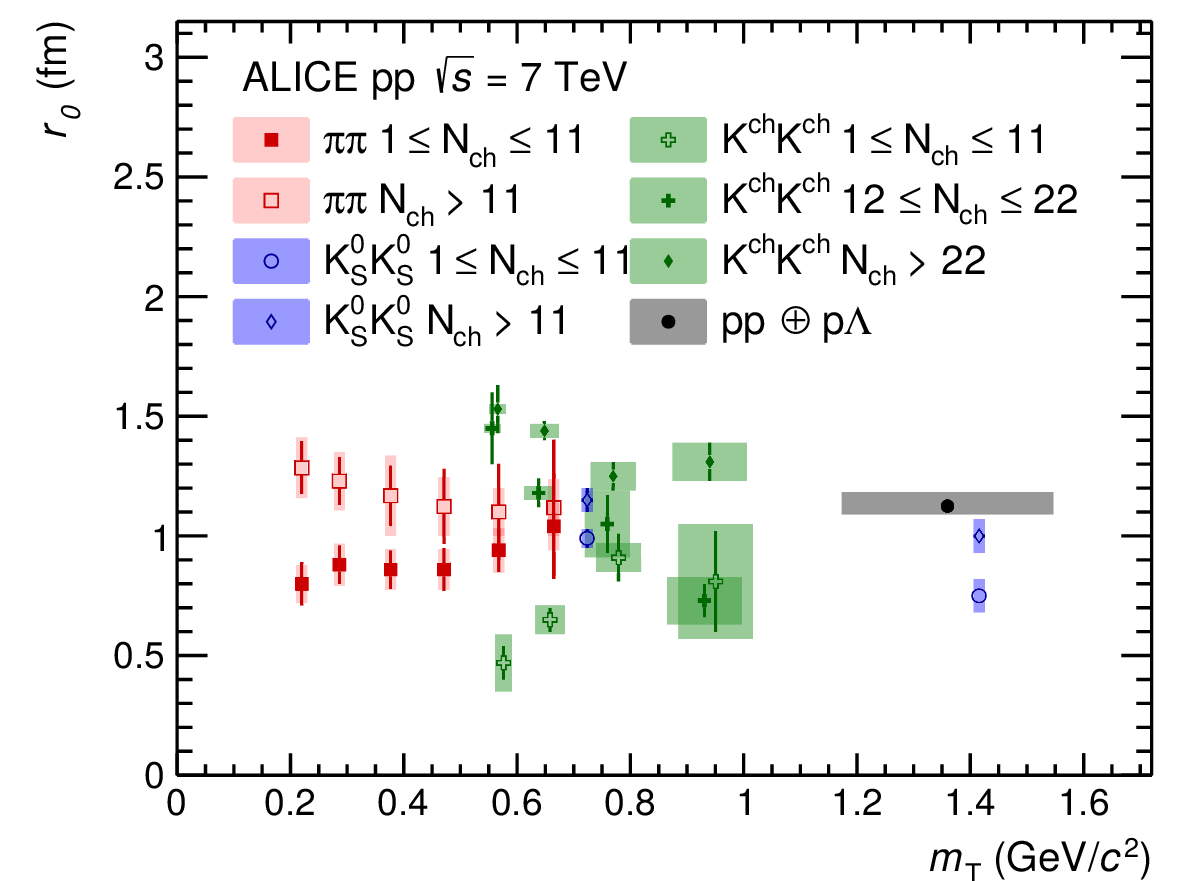}
\caption{\textit{(Color online)} Comparison of radii obtained for different charged particle multiplicity intervals in the \pp collision system at $\sqrt{s}=7$\,TeV~\cite{PhysRevD.84.112004,Abelev:2012sq, 2012151}. The error bars correspond to statistical and the shaded regions to the systematic uncertainties. The black point is the radius obtained in this analysis with \pP, \pL and \LL pairs, while the gray bar corresponds to the range of covered $m_{\mathrm{T}}$ in this analysis.}
\label{fig:RadiusComparison}
\end{figure}

This study is the first measurement with baryon pairs in \pp collisions at $\sqrt{s} = 7\,$TeV, while other femtoscopic analyses were conducted with neutral~\cite{2012151} and charged~\cite{Abelev:2012sq} kaon pairs and charged pion pairs~\cite{PhysRevD.84.112004} with the ALICE experiment. 
The radius obtained from baryon pairs is found to be slightly larger than that measured from meson-meson pairs at comparable transverse mass as shown in Fig.~\ref{fig:RadiusComparison}

\section{Summary}
This paper presents the first femtoscopic measurement of \pP, \pL and \LL pairs
in \pp collisions at $\sqrt{s}=7$\,TeV. No evidence for the presence of mini-jet
background is found and it is demonstrated that this kind of
studies with baryon\mbox{--}baryon and anti-baryon\mbox{--}anti-baryon
pairs are feasible. With a newly developed method to compute the contributions arising from
impurities and weakly decaying resonances to the correlation function from
single particles quantities only, the genuine correlation functions of interest
can be extracted from the signal. These correlation functions contribute with
74\,\% for \pP, 47\,\% for \pL and 30\,\% for \LL to the total signal. A
simultaneous fit of all correlation functions with a femtoscopic model
featuring residual correlations stemming from the above mentioned effects
yields a radius of the particles emitting source of \radiusResult.
For the first time, the Argonne $v_{18}$ $NN$ potential with the $s$
and $p$ waves was used to successfully describe the \pP correlation and in so obtain a solid
benchmark for our investigation. For the case of the \pL correlation function,
the NLO parameter set obtained within the framework of chiral effective field
theory is consistent with the data, but other models are also found to be in agreement with the data.
The present pair data in the \LL channel 
allows us to constrain the available scattering parameter space.
Large effective ranges $d_0$ in combination with negative scattering parameters
lead to unphysical correlations if the Lednick\'{y} model is employed to compute the correlation function.
This also holds true for the the average values published by the STAR
collaboration in Au\mbox{--}Au collisions at $\sqrt{s_{\mathrm{NN}}} = 200$\,GeV, 
that are found to be incompatible with the measurement in \pp collisions within the Lednick\'{y} model.

The larger data sample of the LHC Run 2 and Run 3, where we expect up to a factor ten and 100 more data respectively, will enable us to extend the method also to $\Sigma$, $\Xi$ and $\Omega$ hyperons and thus further constrain the Hyperon\mbox{--}Nucleon interaction.


\newenvironment{acknowledgement}{\relax}{\relax}
\begin{acknowledgement}
\section*{Acknowledgements}

The ALICE Collaboration would like to thank all its engineers and technicians for their invaluable contributions to the construction of the experiment and the CERN accelerator teams for the outstanding performance of the LHC complex.
The ALICE Collaboration gratefully acknowledges the resources and support provided by all Grid centres and the Worldwide LHC Computing Grid (WLCG) collaboration.
The ALICE Collaboration acknowledges the following funding agencies for their support in building and running the ALICE detector:
A. I. Alikhanyan National Science Laboratory (Yerevan Physics Institute) Foundation (ANSL), State Committee of Science and World Federation of Scientists (WFS), Armenia;
Austrian Academy of Sciences and Nationalstiftung f\"{u}r Forschung, Technologie und Entwicklung, Austria;
Ministry of Communications and High Technologies, National Nuclear Research Center, Azerbaijan;
Conselho Nacional de Desenvolvimento Cient\'{\i}fico e Tecnol\'{o}gico (CNPq), Universidade Federal do Rio Grande do Sul (UFRGS), Financiadora de Estudos e Projetos (Finep) and Funda\c{c}\~{a}o de Amparo \`{a} Pesquisa do Estado de S\~{a}o Paulo (FAPESP), Brazil;
Ministry of Science \& Technology of China (MSTC), National Natural Science Foundation of China (NSFC) and Ministry of Education of China (MOEC) , China;
Ministry of Science and Education, Croatia;
Ministry of Education, Youth and Sports of the Czech Republic, Czech Republic;
The Danish Council for Independent Research | Natural Sciences, the Carlsberg Foundation and Danish National Research Foundation (DNRF), Denmark;
Helsinki Institute of Physics (HIP), Finland;
Commissariat \`{a} l'Energie Atomique (CEA) and Institut National de Physique Nucl\'{e}aire et de Physique des Particules (IN2P3) and Centre National de la Recherche Scientifique (CNRS), France;
Bundesministerium f\"{u}r Bildung, Wissenschaft, Forschung und Technologie (BMBF) and GSI Helmholtzzentrum f\"{u}r Schwerionenforschung GmbH, Germany;
General Secretariat for Research and Technology, Ministry of Education, Research and Religions, Greece;
National Research, Development and Innovation Office, Hungary;
Department of Atomic Energy Government of India (DAE), Department of Science and Technology, Government of India (DST), University Grants Commission, Government of India (UGC) and Council of Scientific and Industrial Research (CSIR), India;
Indonesian Institute of Science, Indonesia;
Centro Fermi - Museo Storico della Fisica e Centro Studi e Ricerche Enrico Fermi and Istituto Nazionale di Fisica Nucleare (INFN), Italy;
Institute for Innovative Science and Technology , Nagasaki Institute of Applied Science (IIST), Japan Society for the Promotion of Science (JSPS) KAKENHI and Japanese Ministry of Education, Culture, Sports, Science and Technology (MEXT), Japan;
Consejo Nacional de Ciencia (CONACYT) y Tecnolog\'{i}a, through Fondo de Cooperaci\'{o}n Internacional en Ciencia y Tecnolog\'{i}a (FONCICYT) and Direcci\'{o}n General de Asuntos del Personal Academico (DGAPA), Mexico;
Nederlandse Organisatie voor Wetenschappelijk Onderzoek (NWO), Netherlands;
The Research Council of Norway, Norway;
Commission on Science and Technology for Sustainable Development in the South (COMSATS), Pakistan;
Pontificia Universidad Cat\'{o}lica del Per\'{u}, Peru;
Ministry of Science and Higher Education and National Science Centre, Poland;
Korea Institute of Science and Technology Information and National Research Foundation of Korea (NRF), Republic of Korea;
Ministry of Education and Scientific Research, Institute of Atomic Physics and Romanian National Agency for Science, Technology and Innovation, Romania;
Joint Institute for Nuclear Research (JINR), Ministry of Education and Science of the Russian Federation and National Research Centre Kurchatov Institute, Russia;
Ministry of Education, Science, Research and Sport of the Slovak Republic, Slovakia;
National Research Foundation of South Africa, South Africa;
Centro de Aplicaciones Tecnol\'{o}gicas y Desarrollo Nuclear (CEADEN), Cubaenerg\'{\i}a, Cuba and Centro de Investigaciones Energ\'{e}ticas, Medioambientales y Tecnol\'{o}gicas (CIEMAT), Spain;
Swedish Research Council (VR) and Knut \& Alice Wallenberg Foundation (KAW), Sweden;
European Organization for Nuclear Research, Switzerland;
National Science and Technology Development Agency (NSDTA), Suranaree University of Technology (SUT) and Office of the Higher Education Commission under NRU project of Thailand, Thailand;
Turkish Atomic Energy Agency (TAEK), Turkey;
National Academy of  Sciences of Ukraine, Ukraine;
Science and Technology Facilities Council (STFC), United Kingdom;
National Science Foundation of the United States of America (NSF) and United States Department of Energy, Office of Nuclear Physics (DOE NP), United States of America.    
\end{acknowledgement}

\bibliographystyle{utphys}   
\bibliography{content/Bibliography}

\providecommand{\href}[2]{#2}\begingroup\raggedright\begin{thebibliography}{10}

\bibitem{Pratt:1986cc}
S.~Pratt, ``{Pion Interferometry of Quark-Gluon Plasma},''
\href{http://dx.doi.org/10.1103/PhysRevD.33.1314}{{\em Phys. Rev.} {\bfseries
  D33} (1986) 1314--1327}.

\bibitem{Lisa:2005dd}
M.~A. Lisa, S.~Pratt, R.~Soltz, and U.~Wiedemann, ``{Femtoscopy in relativistic
  heavy ion collisions},''
  \href{http://dx.doi.org/10.1146/annurev.nucl.55.090704.151533}{{\em Ann. Rev.
  Nucl. Part. Sci.} {\bfseries 55} (2005) 357--402},
\href{http://arxiv.org/abs/nucl-ex/0505014}{{\ttfamily arXiv:nucl-ex/0505014
  [nucl-ex]}}.

\bibitem{Henzl:2011dh}
V.~Henzl {\em et~al.}, ``{Angular Dependence in Proton-Proton Correlation
  Functions in Central $^{40}Ca+^{40}Ca$ and $^{48}Ca+^{48}Ca$ Reactions},''
  \href{http://dx.doi.org/10.1103/PhysRevC.85.014606}{{\em Phys. Rev.}
  {\bfseries C85} (2012) 014606},
\href{http://arxiv.org/abs/1108.2552}{{\ttfamily arXiv:1108.2552 [nucl-ex]}}.

\bibitem{Agakishiev:2011zz}
{\bfseries HADES} Collaboration, G.~Agakishiev {\em et~al.}, ``{pp and $\pi\pi$
  intensity interferometry in collisions of Ar + KCl at 1.76A-GeV},''
\href{http://dx.doi.org/10.1140/epja/i2011-11063-x}{{\em Eur. Phys. J.}
  {\bfseries A47} (2011) 63}.

\bibitem{Kotte:2004yv}
{\bfseries FOPI} Collaboration, R.~Kotte {\em et~al.}, ``{Two-proton
  small-angle correlations in central heavy-ion collisions: A Beam-energy and
  system-size dependent study},''
  \href{http://dx.doi.org/10.1140/epja/i2004-10075-y}{{\em Eur. J. Phys.}
  {\bfseries A23} (2005) 271--278},
\href{http://arxiv.org/abs/nucl-ex/0409008}{{\ttfamily arXiv:nucl-ex/0409008
  [nucl-ex]}}.

\bibitem{Aggarwal:2007aa}
{\bfseries WA98} Collaboration, M.~M. Aggarwal {\em et~al.}, ``{Source radii at
  target rapidity from two-proton and two-deuteron correlations in central Pb +
  Pb collisions at 158-A-GeV},''
\href{http://arxiv.org/abs/0709.2477}{{\ttfamily arXiv:0709.2477 [nucl-ex]}}.

\bibitem{Adams:2004yc}
{\bfseries STAR} Collaboration, J.~Adams {\em et~al.}, ``{Pion interferometry
  in Au+Au collisions at
  $\sqrt{{s}_{NN}}=200\phantom{\rule{0.3em}{0ex}}\mathrm{GeV}$},''
  \href{http://dx.doi.org/10.1103/PhysRevC.71.044906}{{\em Phys. Rev. C}
  {\bfseries 71} (Apr, 2005) 044906}.
  \url{https://link.aps.org/doi/10.1103/PhysRevC.71.044906}.

\bibitem{Aamodt:2011mr}
{\bfseries ALICE} Collaboration, K.~Aamodt {\em et~al.}, ``{Two-pion
  Bose-Einstein correlations in central Pb-Pb collisions at $\sqrt{{s}_{NN}} =$
  2.76 TeV},'' \href{http://dx.doi.org/10.1016/j.physletb.2010.12.053}{{\em
  Phys. Lett.} {\bfseries B696} (2011) 328--337},
\href{http://arxiv.org/abs/1012.4035}{{\ttfamily arXiv:1012.4035 [nucl-ex]}}.

\bibitem{Abelev:2014pja}
{\bfseries ALICE} Collaboration, B.~B. Abelev {\em et~al.}, ``{Freeze-out radii
  extracted from three-pion cumulants in pp, p–Pb and Pb–Pb collisions at
  the LHC},'' \href{http://dx.doi.org/10.1016/j.physletb.2014.10.034}{{\em
  Phys. Lett.} {\bfseries B739} (2014) 139--151},
\href{http://arxiv.org/abs/1404.1194}{{\ttfamily arXiv:1404.1194 [nucl-ex]}}.

\bibitem{Adam:2015vna}
{\bfseries ALICE} Collaboration, J.~Adam {\em et~al.}, ``{Centrality dependence
  of pion freeze-out radii in Pb-Pb collisions at $\sqrt{s}_{NN}=$ 2.76 TeV},''
  \href{http://dx.doi.org/10.1103/PhysRevC.93.024905}{{\em Phys. Rev.}
  {\bfseries C93} no.~2, (2016) 024905},
\href{http://arxiv.org/abs/1507.06842}{{\ttfamily arXiv:1507.06842 [nucl-ex]}}.

\bibitem{PhysRevLett.54.302}
C.~B. Chitwood {\em et~al.}, ``{Final-State Interactions between Noncompound
  Light Particles for $^{16}\mathrm{O}$-Induced Reactions on
  $^{197}\mathrm{Au}$ at $\frac{E}{A}=25$ MeV},''
  \href{http://dx.doi.org/10.1103/PhysRevLett.54.302}{{\em Phys. Rev. Lett.}
  {\bfseries 54} (Jan, 1985) 302--305}.
  \url{https://link.aps.org/doi/10.1103/PhysRevLett.54.302}.

\bibitem{Adams:2005ws}
{\bfseries STAR} Collaboration, J.~Adams {\em et~al.},
  ``{Proton-\ensuremath{\Lambda} correlations in central Au+Au collisions at
  $\sqrt{{s}_{\mathit{NN}}}=200$ GeV},''
  \href{http://dx.doi.org/10.1103/PhysRevC.74.064906}{{\em Phys. Rev.}
  {\bfseries C74} (2006) 064906},
\href{http://arxiv.org/abs/nucl-ex/0511003}{{\ttfamily arXiv:nucl-ex/0511003
  [nucl-ex]}}.

\bibitem{Anticic:2011ja}
{\bfseries NA49} Collaboration, T.~Anticic {\em et~al.}, ``{Proton - $\Lambda$
  Correlations in Central Pb+Pb Collisions\ at $\sqrt{s_{NN}} = 17.3$ GeV},''
  \href{http://dx.doi.org/10.1103/PhysRevC.83.054906}{{\em Phys. Rev.}
  {\bfseries C83} (2011) 054906},
\href{http://arxiv.org/abs/1103.3395}{{\ttfamily arXiv:1103.3395 [nucl-ex]}}.

\bibitem{Chung:2002vk}
P.~Chung {\em et~al.}, ``{Comparison of source images for protons, pi-'s and\
  Lambda's in 6-AGeV Au+Au collisions},''
  \href{http://dx.doi.org/10.1103/PhysRevLett.91.162301}{{\em Phys. Rev. Lett.}
  {\bfseries 91} (2003) 162301},
\href{http://arxiv.org/abs/nucl-ex/0212028}{{\ttfamily arXiv:nucl-ex/0212028
  [nucl-ex]}}.

\bibitem{Agakishiev:2010qe}
{\bfseries HADES} Collaboration, G.~Agakishiev {\em et~al.}, ``{Lambda-p
  femtoscopy in collisions of Ar+KCl at 1.76\ AGeV},''
  \href{http://dx.doi.org/10.1103/PhysRevC.82.021901}{{\em Phys. Rev.}
  {\bfseries C82} (2010) 021901},
\href{http://arxiv.org/abs/1004.2328}{{\ttfamily arXiv:1004.2328 [nucl-ex]}}.

\bibitem{Adamczyk:2014vca}
{\bfseries STAR} Collaboration, L.~Adamczyk {\em et~al.}, ``{$\Lambda\Lambda$
  Correlation Function in Au+Au collisions at $\sqrt{s_{NN}}=$ 200 GeV},''
  \href{http://dx.doi.org/10.1103/PhysRevLett.114.022301}{{\em Phys. Rev.
  Lett.} {\bfseries 114} no.~2, (2015) 022301},
\href{http://arxiv.org/abs/1408.4360}{{\ttfamily arXiv:1408.4360 [nucl-ex]}}.

\bibitem{Adamczyk:2015hza}
{\bfseries STAR} Collaboration, L.~Adamczyk {\em et~al.}, ``{Measurement of
  interaction between antiprotons},''
  \href{http://dx.doi.org/10.1038/nature15724}{{\em Nature} (2015) },
\href{http://arxiv.org/abs/1507.07158}{{\ttfamily arXiv:1507.07158 [nucl-ex]}}.

\bibitem{Shapoval:2014yha}
V.~M. Shapoval, B.~Erazmus, R.~Lednicky, and {\relax Yu}.~M. Sinyukov,
  ``{Extracting $p\Lambda$ scattering lengths from heavy ion collisions},''
  \href{http://dx.doi.org/10.1103/PhysRevC.92.034910}{{\em Phys. Rev.}
  {\bfseries C92} no.~3, (2015) 034910},
\href{http://arxiv.org/abs/1405.3594}{{\ttfamily arXiv:1405.3594 [nucl-th]}}.

\bibitem{Kisiel:2014mma}
A.~Kisiel, H.~Zbroszczyk, and M.~Szymanski, ``{Extracting baryon-antibaryon
  strong interaction potentials from p$\bar{\Lambda}$ femtoscopic correlation
  functions},'' \href{http://dx.doi.org/10.1103/PhysRevC.89.054916}{{\em Phys.
  Rev.} {\bfseries C89} no.~5, (2014) 054916},
\href{http://arxiv.org/abs/1403.0433}{{\ttfamily arXiv:1403.0433 [nucl-th]}}.

\bibitem{Weise:2009ny}
W.~Weise, ``{Low-energy QCD and hadronic structure},''
  \href{http://dx.doi.org/10.1016/j.nuclphysa.2009.05.020}{{\em Nucl. Phys.}
  {\bfseries A827} (2009) 66C--76C},
  \href{http://arxiv.org/abs/0905.4898}{{\ttfamily arXiv:0905.4898 [nucl-th]}}.
[,66(2009)].

\bibitem{SechiZorn:1969hk}
B.~Sechi-Zorn, B.~Kehoe, J.~Twitty, and R.~A. Burnstein, ``{Low-energy
  $\Lambda$-Proton elastic scattering},''
\href{http://dx.doi.org/10.1103/PhysRev.175.1735}{{\em Phys. Rev.} {\bfseries
  175} (1968) 1735--1740}.

\bibitem{Eisele:1971mk}
F.~Eisele, H.~Filthuth, W.~Foehlisch, V.~Hepp, and G.~Zech, ``{Elastic
  $\Sigma^{\pm}$p scattering at low energies},''
\href{http://dx.doi.org/10.1016/0370-2693(71)90053-0}{{\em Phys. Lett.}
  {\bfseries B37} (1971) 204--206}.

\bibitem{Alexander:1969cx}
G.~Alexander {\em et~al.}, ``{Study of the $\Lambda-n$ system in low-energy
  $\Lambda-p$ elastic scattering},''
\href{http://dx.doi.org/10.1103/PhysRev.173.1452}{{\em Phys. Rev.} {\bfseries
  173} (1968) 1452--1460}.

\bibitem{Nagels:2015lfa}
M.~M. Nagels, T.~A. Rijken, and Y.~Yamamoto, ``{Extended-soft-core
  Baryon-Baryon Model Esc08 II. Hyperon-Nucleon Interactions},''
\href{http://arxiv.org/abs/1501.06636}{{\ttfamily arXiv:1501.06636 [nucl-th]}}.

\bibitem{POLINDER2006244}
H.~Polinder, J.~Haidenbauer, and U.-G. Mei{\ss}ner, ``{Hyperon-nucleon
  interactions - a chiral effective field theory approach},''
  \href{http://dx.doi.org/https://doi.org/10.1016/j.nuclphysa.2006.09.006}{{\em
  Nuclear Physics A} {\bfseries 779} (2006) 244 -- 266}.
  \url{http://www.sciencedirect.com/science/article/pii/S0375947406006312}.

\bibitem{Haidenbauer:2013oca}
J.~Haidenbauer, S.~Petschauer, N.~Kaiser, U.~G. Meissner, A.~Nogga, and
  W.~Weise, ``{Hyperon-nucleon interaction at next-to-leading order in chiral
  effective field theory},''
  \href{http://dx.doi.org/10.1016/j.nuclphysa.2013.06.008}{{\em Nucl. Phys.}
  {\bfseries A915} (2013) 24--58},
\href{http://arxiv.org/abs/1304.5339}{{\ttfamily arXiv:1304.5339 [nucl-th]}}.

\bibitem{Hashimoto:2006aw}
O.~Hashimoto and H.~Tamura, ``{Spectroscopy of $\Lambda$ hypernuclei},''
\href{http://dx.doi.org/10.1016/j.ppnp.2005.07.001}{{\em Prog. Part. Nucl.
  Phys.} {\bfseries 57} (2006) 564--653}.

\bibitem{Hayano:1988pn}
R.~S. Hayano {\em et~al.}, ``{Observation of a Bound State of $^{4}$He
  ($\Sigma$) Hypernucleus},''
\href{http://dx.doi.org/10.1016/0370-2693(89)90675-8}{{\em Phys. Lett.}
  {\bfseries B231} (1989) 355--358}.

\bibitem{PhysRevLett.80.1605}
T.~Nagae {\em et~al.}, ``{Observation of a
  ${}_{\ensuremath{\Sigma}}^{4}\mathrm{He}$ Bound State in the
  $^{4}He({K}^{\ensuremath{-}},{\ensuremath{\pi}}^{\ensuremath{-}})$ Reaction
  at $600 MeV/\mathit{c}$},''
  \href{http://dx.doi.org/10.1103/PhysRevLett.80.1605}{{\em Phys. Rev. Lett.}
  {\bfseries 80} (Feb, 1998) 1605--1609}.
  \url{https://link.aps.org/doi/10.1103/PhysRevLett.80.1605}.

\bibitem{Nemura:2017vjc}
H.~Nemura {\em et~al.}, ``{Baryon interactions from lattice QCD with physical
  masses --- strangeness $S=-1$ sector ---},'' in {\em {35th International
  Symposium on Lattice Field Theory (Lattice 2017) Granada, Spain, June 18-24,
  2017}}.
\newblock 2017.
\newblock \href{http://arxiv.org/abs/1711.07003}{{\ttfamily arXiv:1711.07003
  [hep-lat]}}.
\newblock
\url{http://inspirehep.net/record/1637203/files/arXiv:1711.07003.pdf}.
\newblock

\bibitem{Nakazawa:15}
K.~Nakazawa {\em et~al.}, ``{The first evidence of a deeply bound state of
  Xi-14N system},'' \href{http://dx.doi.org/10.1093/ptep/ptv008}{{\em Progress
  of Theoretical and Experimental Physics} {\bfseries 2015} no.~3, (2015)
  033D02}. \url{+ http://dx.doi.org/10.1093/ptep/ptv008}.

\bibitem{Nagae:2017slp}
T.~Nagae {\em et~al.}, ``{Search For A $\Xi$ Bound State In The
  $^{12}$C($K^-$,$K^+$)$X$ Reaction At 1.8 Gev/c in J-PARC},''
{\em PoS} {\bfseries INPC2016} (2017) 038.

\bibitem{Hatsuda:2017uxk}
T.~Hatsuda, K.~Morita, A.~Ohnishi, and K.~Sasaki, ``{$p\Xi^- $ Correlation in
  Relativistic Heavy Ion Collisions with Nucleon-Hyperon Interaction from
  Lattice QCD},'' \href{http://dx.doi.org/10.1016/j.nuclphysa.2017.04.041}{{\em
  Nucl. Phys.} {\bfseries A967} (2017) 856--859},
\href{http://arxiv.org/abs/1704.05225}{{\ttfamily arXiv:1704.05225 [nucl-th]}}.

\bibitem{PhysRevLett.83.3138}
F.~Wang and S.~Pratt, ``Lambda-proton correlations in relativistic heavy ion
  collisions,'' \href{http://dx.doi.org/10.1103/PhysRevLett.83.3138}{{\em Phys.
  Rev. Lett.} {\bfseries 83} (Oct, 1999) 3138--3141}.
  \url{https://link.aps.org/doi/10.1103/PhysRevLett.83.3138}.

\bibitem{PhysRevLett.38.195}
R.~L. Jaffe, ``Perhaps a stable dihyperon,''
  \href{http://dx.doi.org/10.1103/PhysRevLett.38.195}{{\em Phys. Rev. Lett.}
  {\bfseries 38} (Jan, 1977) 195--198}.
  \url{https://link.aps.org/doi/10.1103/PhysRevLett.38.195}.

\bibitem{Takahashi:2001nm}
H.~Takahashi {\em et~al.}, ``{Observation of a
  $_{\ensuremath{\Lambda}\ensuremath{\Lambda}}^{6}$He double hypernucleus},''
\href{http://dx.doi.org/10.1103/PhysRevLett.87.212502}{{\em Phys. Rev. Lett.}
  {\bfseries 87} (2001) 212502}.

\bibitem{Sasaki:2017ysy}
K.~Sasaki {\em et~al.}, ``{Baryon interactions from lattice QCD with physical
  masses -- $S=-2$ sector --},'' {\em PoS} {\bfseries LATTICE2016} (2017) 116,
\href{http://arxiv.org/abs/1702.06241}{{\ttfamily arXiv:1702.06241 [hep-lat]}}.

\bibitem{PhysRevC.91.024916}
K.~Morita, T.~Furumoto, and A.~Ohnishi,
  ``{$\ensuremath{\Lambda}\ensuremath{\Lambda}$ interaction from relativistic
  heavy-ion collisions},''
  \href{http://dx.doi.org/10.1103/PhysRevC.91.024916}{{\em Phys. Rev. C}
  {\bfseries 91} (Feb, 2015) 024916}.
  \url{https://link.aps.org/doi/10.1103/PhysRevC.91.024916}.

\bibitem{Petschauer:2015nea}
S.~Petschauer, J.~Haidenbauer, N.~Kaiser, U.-G. Mei{\ss}ner, and W.~Weise,
  ``{Hyperons in nuclear matter from SU(3) chiral effective field theory},''
  \href{http://dx.doi.org/10.1140/epja/i2016-16015-4}{{\em The European
  Physical Journal A} {\bfseries 52} no.~1, (Jan, 2016) 15}.
  \url{https://doi.org/10.1140/epja/i2016-16015-4}.

\bibitem{Schulze:2006vw}
H.~J. Schulze, A.~Polls, A.~Ramos, and I.~Vidana, ``{Maximum mass of neutron
  stars},''
\href{http://dx.doi.org/10.1103/PhysRevC.73.058801}{{\em Phys. Rev.} {\bfseries
  C73} (2006) 058801}.

\bibitem{Weissenborn:2011kb}
S.~Weissenborn, D.~Chatterjee, and J.~Schaffner-Bielich, ``{Hyperons and
  massive neutron stars: the role of hyperon potentials},''
  \href{http://dx.doi.org/10.1016/j.nuclphysa.2012.02.012}{{\em Nucl. Phys.}
  {\bfseries A881} (2012) 62--77},
\href{http://arxiv.org/abs/1111.6049}{{\ttfamily arXiv:1111.6049
  [astro-ph.HE]}}.

\bibitem{Weissenborn:2011ut}
S.~Weissenborn, D.~Chatterjee, and J.~Schaffner-Bielich, ``{Hyperons and
  massive neutron stars: vector repulsion and SU(3) symmetry},''
  \href{http://dx.doi.org/10.1103/PhysRevC.85.065802,
  10.1103/PhysRevC.90.019904}{{\em Phys. Rev.} {\bfseries C85} no.~6, (2012)
  065802}, \href{http://arxiv.org/abs/1112.0234}{{\ttfamily arXiv:1112.0234
  [astro-ph.HE]}}.
[Erratum: Phys. Rev.C90,no.1,019904(2014)].

\bibitem{Djapo:2008au}
H.~Djapo, B.-J. Schaefer, and J.~Wambach, ``{On the appearance of hyperons in
  neutron stars},'' \href{http://dx.doi.org/10.1103/PhysRevC.81.035803}{{\em
  Phys. Rev.} {\bfseries C81} (2010) 035803},
\href{http://arxiv.org/abs/0811.2939}{{\ttfamily arXiv:0811.2939 [nucl-th]}}.

\bibitem{Demorest:2010bx}
P.~Demorest, T.~Pennucci, S.~Ransom, M.~Roberts, and J.~Hessels, ``{Shapiro
  Delay Measurement of A Two Solar Mass Neutron Star},''
  \href{http://dx.doi.org/10.1038/nature09466}{{\em Nature} {\bfseries 467}
  (2010) 1081--1083},
\href{http://arxiv.org/abs/1010.5788}{{\ttfamily arXiv:1010.5788
  [astro-ph.HE]}}.

\bibitem{Antoniadis:2013pzd}
J.~Antoniadis {\em et~al.}, ``{A Massive Pulsar in a Compact Relativistic
  Binary},'' \href{http://dx.doi.org/10.1126/science.1233232}{{\em Science}
  {\bfseries 340} (2013) 6131},
\href{http://arxiv.org/abs/1304.6875}{{\ttfamily arXiv:1304.6875
  [astro-ph.HE]}}.

\bibitem{Yamamoto:2013ada}
Y.~Yamamoto, T.~Furumoto, N.~Yasutake, and T.~A. Rijken, ``{Multi-pomeron
  repulsion and the Neutron-star mass},''
  \href{http://dx.doi.org/10.1103/PhysRevC.88.022801}{{\em Phys. Rev.}
  {\bfseries C88} no.~2, (2013) 022801},
\href{http://arxiv.org/abs/1308.2130}{{\ttfamily arXiv:1308.2130 [nucl-th]}}.

\bibitem{Yamamoto:2014jga}
Y.~Yamamoto, T.~Furumoto, N.~Yasutake, and T.~A. Rijken, ``{Hyperon mixing and
  universal many-body repulsion in neutron stars},''
  \href{http://dx.doi.org/10.1103/PhysRevC.90.045805}{{\em Phys. Rev.}
  {\bfseries C90} (2014) 045805},
\href{http://arxiv.org/abs/1406.4332}{{\ttfamily arXiv:1406.4332 [nucl-th]}}.

\bibitem{Oertel:2016bki}
M.~Oertel, M.~Hempel, T.~Kl{\"a}hn, and S.~Typel, ``{Equations of state for
  supernovae and compact stars},''
  \href{http://dx.doi.org/10.1103/RevModPhys.89.015007}{{\em Rev. Mod. Phys.}
  {\bfseries 89} no.~1, (2017) 015007},
  \href{http://arxiv.org/abs/1610.03361}{{\ttfamily arXiv:1610.03361}}.

\bibitem{Lonardoni:2014bwa}
D.~Lonardoni, A.~Lovato, S.~Gandolfi, and F.~Pederiva, ``{Hyperon Puzzle: Hints
  from Quantum Monte Carlo Calculations},''
  \href{http://dx.doi.org/10.1103/PhysRevLett.114.092301}{{\em Phys. Rev.
  Lett.} {\bfseries 114} no.~9, (2015) 092301},
\href{http://arxiv.org/abs/1407.4448}{{\ttfamily arXiv:1407.4448 [nucl-th]}}.

\bibitem{PhysRevD.84.112004}
{\bfseries ALICE} Collaboration, K.~Aamodt {\em et~al.}, ``{Femtoscopy of $pp$
  collisions at $\sqrt{s}=0.9$ and 7 TeV at the LHC with two-pion Bose-Einstein
  correlations},'' \href{http://dx.doi.org/10.1103/PhysRevD.84.112004}{{\em
  Phys. Rev. D} {\bfseries 84} (Dec, 2011) 112004}.
  \url{https://link.aps.org/doi/10.1103/PhysRevD.84.112004}.

\bibitem{Abelev:2012sq}
{\bfseries ALICE} Collaboration, B.~Abelev {\em et~al.}, ``{Charged kaon
  femtoscopic correlations in $pp$ collisions at $\sqrt{s}=7$ TeV},''
  \href{http://dx.doi.org/10.1103/PhysRevD.87.052016}{{\em Phys. Rev.}
  {\bfseries D87} no.~5, (2013) 052016},
\href{http://arxiv.org/abs/1212.5958}{{\ttfamily arXiv:1212.5958 [hep-ex]}}.

\bibitem{PhysRevC.51.38}
R.~B. Wiringa, V.~G.~J. Stoks, and R.~Schiavilla, ``Accurate nucleon-nucleon
  potential with charge-independence breaking,''
  \href{http://dx.doi.org/10.1103/PhysRevC.51.38}{{\em Phys. Rev. C} {\bfseries
  51} (Jan, 1995) 38--51}.
  \url{https://link.aps.org/doi/10.1103/PhysRevC.51.38}.

\bibitem{CATS}
D.~L. Mihaylov, V.~M. Sarti, O.~W. Arnold, L.~Fabbietti, B.~Hohlweger, and
  A.~M. Mathis, ``{A femtoscopic Correlation Analysis Tool using the
  Schr\"odinger equation (CATS)},''
  \href{http://dx.doi.org/10.1140/epjc/s10052-018-5859-0}{{\em Eur. Phys. J.}
  {\bfseries C78} no.~5, (2018) 394},
\href{http://arxiv.org/abs/1802.08481}{{\ttfamily arXiv:1802.08481 [hep-ph]}}.

\bibitem{1748-0221-3-08-S08002}
{\bfseries ALICE} Collaboration, K.~Aamodt {\em et~al.}, ``{The ALICE
  experiment at the CERN LHC},'' {\em Journal of Instrumentation} {\bfseries 3}
  no.~08, (2008) S08002. \url{http://stacks.iop.org/1748-0221/3/i=08/a=S08002}.

\bibitem{Abelev:2014ffa}
{\bfseries ALICE} Collaboration, B.~B. Abelev {\em et~al.}, ``{Performance of
  the ALICE Experiment at the CERN LHC},''
  \href{http://dx.doi.org/10.1142/S0217751X14300440}{{\em Int. J. Mod. Phys.}
  {\bfseries A29} (2014) 1430044},
\href{http://arxiv.org/abs/1402.4476}{{\ttfamily arXiv:1402.4476 [nucl-ex]}}.

\bibitem{2010TPCNIMA}
J.~{Alme} {\em et~al.}, ``{The ALICE TPC, a large 3-dimensional tracking device
  with fast readout for ultra-high multiplicity events},''
  \href{http://dx.doi.org/10.1016/j.nima.2010.04.042}{{\em Nuclear Instruments
  and Methods in Physics Research A} {\bfseries 622} (Oct., 2010) 316--367},
  \href{http://arxiv.org/abs/1001.1950}{{\ttfamily arXiv:1001.1950
  [physics.ins-det]}}.

\bibitem{Akindinov2013}
A.~Akindinov {\em et~al.}, ``{Performance of the ALICE Time-Of-Flight detector
  at the LHC},'' \href{http://dx.doi.org/10.1140/epjp/i2013-13044-x}{{\em The
  European Physical Journal Plus} {\bfseries 128} no.~4, (Apr, 2013) 44}.
  \url{https://doi.org/10.1140/epjp/i2013-13044-x}.

\bibitem{Adam2017}
{\bfseries ALICE} Collaboration, J.~Adam {\em et~al.}, ``{Insight into particle
  production mechanisms via angular correlations of identified particles in pp
  collisions at $\sqrt{s} = 7$ TeV},''
  \href{http://dx.doi.org/10.1140/epjc/s10052-017-5129-6}{{\em The European
  Physical Journal C} {\bfseries 77} no.~8, (Aug, 2017) 569}.
  \url{https://doi.org/10.1140/epjc/s10052-017-5129-6}.

\bibitem{Patrignani:2016xqp}
{\bfseries Particle Data Group} Collaboration, C.~Patrignani {\em et~al.},
  ``{Review of Particle Physics},''
\href{http://dx.doi.org/10.1088/1674-1137/40/10/100001}{{\em Chin. Phys.}
  {\bfseries C40} no.~10, (2016) 100001}.

\bibitem{0954-3899-32-10-001}
{\bfseries ALICE} Collaboration, B.~Alessandro {\em et~al.}, ``{ALICE: Physics
  Performance Report, Volume II},'' {\em Journal of Physics G: Nuclear and
  Particle Physics} {\bfseries 32} no.~10, (2006) 1295.
  \url{http://stacks.iop.org/0954-3899/32/i=10/a=001}.

\bibitem{1126-6708-2006-05-026}
T.~Sj\"{o}strand, S.~Mrenna, and P.~Skands, ``Pythia 6.4 physics and manual,''
  {\em Journal of High Energy Physics} {\bfseries 2006} no.~05, (2006) 026.
  \url{http://stacks.iop.org/1126-6708/2006/i=05/a=026}.

\bibitem{PhysRevD.82.074018}
P.~Z. Skands, ``Tuning monte carlo generators: The perugia tunes,''
  \href{http://dx.doi.org/10.1103/PhysRevD.82.074018}{{\em Phys. Rev. D}
  {\bfseries 82} (Oct, 2010) 074018}.
  \url{https://link.aps.org/doi/10.1103/PhysRevD.82.074018}.

\bibitem{Abbas2013}
E.~Abbas {\em et~al.}, ``{Mid-rapidity anti-baryon to baryon ratios in pp
  collisions at $\sqrt{s}= 0.9$, $2.76$ and $7$\,TeV measured by ALICE},''
  \href{http://dx.doi.org/10.1140/epjc/s10052-013-2496-5}{{\em The European
  Physical Journal C} {\bfseries 73} no.~7, (Jul, 2013) 2496}.
  \url{https://doi.org/10.1140/epjc/s10052-013-2496-5}.

\bibitem{Adamczewski-Musch:2017gjr}
{\bfseries HADES} Collaboration, J.~Adamczewski-Musch {\em et~al.},
  ``{$\Sigma^0$ production in proton nucleus collisions near threshold},''
\href{http://arxiv.org/abs/1711.05559}{{\ttfamily arXiv:1711.05559 [nucl-ex]}}.

\bibitem{200079}
{\bfseries L3} Collaboration, M.~Acciarri {\em et~al.}, ``{Inclusive $\Sigma^+$
  and $\Sigma^0$ production in hadronic Z decays},''
  \href{http://dx.doi.org/https://doi.org/10.1016/S0370-2693(00)00369-5}{{\em
  Physics Letters B} {\bfseries 479} no.~1, (2000) 79 -- 88}.
  \url{http://www.sciencedirect.com/science/article/pii/S0370269300003695}.

\bibitem{ACCIARRI1994223}
{\bfseries L3} Collaboration, M.~Acciarri {\em et~al.}, ``{Measurement of
  inclusive production of neutral hadrons from Z decays},''
  \href{http://dx.doi.org/https://doi.org/10.1016/0370-2693(94)90453-7}{{\em
  Physics Letters B} {\bfseries 328} no.~1, (1994) 223 -- 233}.
  \url{http://www.sciencedirect.com/science/article/pii/0370269394904537}.

\bibitem{0954-3899-31-6-072}
{\bfseries STAR} Collaboration, G.~V. Buren, ``{The $\Sigma ^{0}/\Lambda$ ratio
  in high energy nuclear collisions},'' {\em Journal of Physics G: Nuclear and
  Particle Physics} {\bfseries 31} no.~6, (2005) S1127.
  \url{http://stacks.iop.org/0954-3899/31/i=6/a=072}.

\bibitem{Becattini:2009ee}
F.~Becattini, P.~Castorina, A.~Milov, and H.~Satz, ``{Predictions of hadron
  abundances in pp collisions at the LHC},''
  \href{http://dx.doi.org/10.1088/0954-3899/38/2/025002}{{\em J. Phys.}
  {\bfseries G38} (2011) 025002},
\href{http://arxiv.org/abs/0912.2855}{{\ttfamily arXiv:0912.2855 [hep-ph]}}.

\bibitem{PhysRevC.91.034906}
{\bfseries ALICE} Collaboration, J.~Adam {\em et~al.}, ``{Two-pion femtoscopy
  in $p$-Pb collisions at $\sqrt{{s}_{NN}}=5.02$ TeV},''
  \href{http://dx.doi.org/10.1103/PhysRevC.91.034906}{{\em Phys. Rev. C}
  {\bfseries 91} (Mar, 2015) 034906}.
  \url{https://link.aps.org/doi/10.1103/PhysRevC.91.034906}.

\bibitem{Bock:2011}
N.~Bock, {\em Femtoscopy of proton-proton collisions in the ALICE experiment}.
\newblock PhD thesis, Ohio State University, 2011.

\bibitem{KOONIN197743}
S.~E. Koonin, ``Proton pictures of high-energy nuclear collisions,''
  \href{http://dx.doi.org/https://doi.org/10.1016/0370-2693(77)90340-9}{{\em
  Physics Letters B} {\bfseries 70} no.~1, (1977) 43 -- 47}.
  \url{http://www.sciencedirect.com/science/article/pii/0370269377903409}.

\bibitem{Ledn}
R.~Lednick\'{y} and V.~Lyuboshits, ``{Final State Interaction Effect on Pairing
  Correlations Between Particles with Small Relative Momenta},'' {\em Sov. J.
  Nucl. Phys.} {\bfseries 35} (1982) 770.

\bibitem{PhysRevC.60.067901}
F.~Wang, ``{Residual correlation in two-proton interferometry from
  $\ensuremath{\Lambda}$-proton strong interactions},''
  \href{http://dx.doi.org/10.1103/PhysRevC.60.067901}{{\em Phys. Rev. C}
  {\bfseries 60} (Nov, 1999) 067901}.
  \url{https://link.aps.org/doi/10.1103/PhysRevC.60.067901}.

\bibitem{Stavinskiy:2007wb}
A.~Stavinskiy, K.~Mikhailov, B.~Erazmus, and R.~Lednicky, ``{Residual
  correlations between decay products of $\pi^{0}\pi^{0}$ and $p\Sigma^{0}$
  systems},''
\href{http://arxiv.org/abs/0704.3290}{{\ttfamily arXiv:0704.3290 [nucl-th]}}.

\bibitem{PhysRevC.29.684}
A.~R. Bodmer, Q.~N. Usmani, and J.~Carlson, ``Binding energies of hypernuclei
  and three-body $\ensuremath{\Lambda}\mathrm{NN}$ forces,''
  \href{http://dx.doi.org/10.1103/PhysRevC.29.684}{{\em Phys. Rev. C}
  {\bfseries 29} (Feb, 1984) 684--687}.
  \url{https://link.aps.org/doi/10.1103/PhysRevC.29.684}.

\bibitem{2012151}
{\bfseries ALICE} Collaboration, B.~Abelev {\em et~al.},
  ``{K$_{s}^{0}$K$_{s}^{0}$ correlations in pp collisions at $\sqrt{s}=7$\,TeV
  from the LHC ALICE experiment},''
  \href{http://dx.doi.org/https://doi.org/10.1016/j.physletb.2012.09.013}{{\em
  Physics Letters B} {\bfseries 717} no.~1, (2012) 151 -- 161}.
  \url{http://www.sciencedirect.com/science/article/pii/S0370269312009574}.

\bibitem{ND}
M.~M. Nagels, T.~A. Rijken, and J.~J. de~Swart, ``{Baryon-baryon scattering in
  a one-boson-exchange-potential approach. II. Hyperon-nucleon scattering},''
  \href{http://dx.doi.org/10.1103/PhysRevD.15.2547}{{\em Phys. Rev. D}
  {\bfseries 15} (May, 1977) 2547--2564}.
  \url{https://link.aps.org/doi/10.1103/PhysRevD.15.2547}.

\bibitem{NF}
M.~M. Nagels, T.~A. Rijken, and J.~J. de~Swart, ``{Baryon-baryon scattering in
  a one-boson-exchange-potential approach. III. A nucleon-nucleon and
  hyperon-nucleon analysis including contributions of a nonet of scalar
  mesons},'' \href{http://dx.doi.org/10.1103/PhysRevD.20.1633}{{\em Phys. Rev.
  D} {\bfseries 20} (Oct, 1979) 1633--1645}.
  \url{https://link.aps.org/doi/10.1103/PhysRevD.20.1633}.

\bibitem{NSC89}
P.~M.~M. Maessen, T.~A. Rijken, and J.~J. de~Swart, ``{Soft-core baryon-baryon
  one-boson-exchange models. II. Hyperon-nucleon potential},''
  \href{http://dx.doi.org/10.1103/PhysRevC.40.2226}{{\em Phys. Rev. C}
  {\bfseries 40} (Nov, 1989) 2226--2245}.
  \url{https://link.aps.org/doi/10.1103/PhysRevC.40.2226}.

\bibitem{NSC97}
T.~A. Rijken, V.~G.~J. Stoks, and Y.~Yamamoto, ``Soft-core hyperon-nucleon
  potentials,'' \href{http://dx.doi.org/10.1103/PhysRevC.59.21}{{\em Phys. Rev.
  C} {\bfseries 59} (Jan, 1999) 21--40}.
  \url{https://link.aps.org/doi/10.1103/PhysRevC.59.21}.

\bibitem{ESC08}
T.~A. Rijken, M.~M. Nagels, and Y.~Yamamoto, ``Baryon-baryon interactions-
  nijmegen extended-soft-core models -,''
  \href{http://dx.doi.org/10.1143/PTPS.185.14}{{\em Progress of Theoretical
  Physics Supplement} {\bfseries 185} (2010) 14--71}.

\bibitem{JulichA}
B.~Holzenkamp, K.~Holinde, and J.~Speth, ``A meson exchange model for the
  hyperon-nucleon interaction,''
  \href{http://dx.doi.org/https://doi.org/10.1016/0375-9474(89)90223-6}{{\em
  Nuclear Physics A} {\bfseries 500} no.~3, (1989) 485 -- 528}.
  \url{http://www.sciencedirect.com/science/article/pii/0375947489902236}.

\bibitem{JulichJ}
J.~Haidenbauer and U.-G. Mei\ss{}ner, ``J\"ulich hyperon-nucleon model
  revisited,'' \href{http://dx.doi.org/10.1103/PhysRevC.72.044005}{{\em Phys.
  Rev. C} {\bfseries 72} (Oct, 2005) 044005}.
  \url{https://link.aps.org/doi/10.1103/PhysRevC.72.044005}.

\bibitem{Ehime1}
T.~Ueda {\em et~al.}, ``{$\ensuremath{\Lambda}\mathrm{N}$ and
  $\ensuremath{\Lambda\Lambda}$ Interactions in an OBE Model and
  Hypernuclei},'' \href{http://dx.doi.org/10.1143/PTP.99.891}{{\em Progress of
  Theoretical Physics} {\bfseries 99} no.~5, (1998) 891--896}.

\bibitem{Ehime2}
K.~Tominaga {\em et~al.}, ``{A one-boson-exchange potential for
  $\ensuremath{\Lambda}\mathrm{N}$, $\ensuremath{\Lambda\Lambda}$ and
  $\ensuremath{\Xi}\mathrm{N}$ systems and hypernuclei},''
  \href{http://dx.doi.org/https://doi.org/10.1016/S0375-9474(98)00485-0}{{\em
  Nuclear Physics A} {\bfseries 642} no.~3, (1998) 483 -- 505}.
  \url{http://www.sciencedirect.com/science/article/pii/S0375947498004850}.

\bibitem{DANYSZ1963121}
M.~Danysz {\em et~al.}, ``The identification of a double hyperfragment,''
  \href{http://dx.doi.org/https://doi.org/10.1016/0029-5582(63)90080-4}{{\em
  Nuclear Physics} {\bfseries 49} (1963) 121 -- 132}.
  \url{http://www.sciencedirect.com/science/article/pii/0029558263900804}.

\bibitem{fss1}
Y.~Fujiwara, Y.~Suzuki, and C.~Nakamoto, ``{Baryon-baryon interactions in the
  SU6 quark model and their applications to light nuclear systems},''
  \href{http://dx.doi.org/https://doi.org/10.1016/j.ppnp.2006.08.001}{{\em
  Progress in Particle and Nuclear Physics} {\bfseries 58} no.~2, (2007) 439 --
  520}.
  \url{http://www.sciencedirect.com/science/article/pii/S0146641006000718}.

\bibitem{fss2}
Y.~Fujiwara, M.~Kohno, C.~Nakamoto, and Y.~Suzuki, ``{Interactions between
  octet baryons in the ${\mathrm{SU}}_{6}$ quark model},''
  \href{http://dx.doi.org/10.1103/PhysRevC.64.054001}{{\em Phys. Rev. C}
  {\bfseries 64} (Sep, 2001) 054001}.
  \url{https://link.aps.org/doi/10.1103/PhysRevC.64.054001}.

\bibitem{FG}
I.~Filikhin and A.~Gal, ``{Faddeev-Yakubovsky calculations for light
  $\ensuremath{\Lambda\Lambda}$ hypernuclei},''
  \href{http://dx.doi.org/https://doi.org/10.1016/S0375-9474(02)01008-4}{{\em
  Nuclear Physics A} {\bfseries 707} no.~3, (2002) 491 -- 509}.
  \url{http://www.sciencedirect.com/science/article/pii/S0375947402010084}.

\bibitem{HKMYY}
E.~Hiyama, M.~Kamimura, T.~Motoba, T.~Yamada, and Y.~Yamamoto, ``{Four-body
  cluster structure of $A=7-10$ double-\ensuremath{\Lambda} hypernuclei},''
  \href{http://dx.doi.org/10.1103/PhysRevC.66.024007}{{\em Phys. Rev. C}
  {\bfseries 66} (Aug, 2002) 024007}.
  \url{https://link.aps.org/doi/10.1103/PhysRevC.66.024007}.

\end{thebibliography}\endgroup


\providecommand{\href}[2]{#2}\begingroup\raggedright\endgroup

\newpage
\appendix
\section{Derivation of the \ensuremath{\mathbf{\lambda}} parameters}
\label{Seq:AppLambdaParam}
Let 'X' be a specific particle type and $X$ is the number of particles of that species. For each particle different subsets 
$X_i$ are defined, each representing a unique origin of the particle, where $i=0$ corresponds to the case of a primary particle, the rest are either 
particles originating from feed-down or misidentification. 
In particular indexes $1\leq i\leq N_F$ should be associated with feed-down contributions and 
$N_F+1\leq i\leq N_F+N_M$ should be associated with impurities, where $N_F$ is the number of feed-down channels and $N_M$ the number of impurity channels. 
In the present work we assume that all impurity channels contribute with a flat distribution to the total correlation, therefore we do not study 
differentially the origin of the impurities and combine them in a single channel, i.e. $N_M=1$. 
Further we define 
\begin{equation}
 X_F = \sum_{i-1}^{N_F} X_i,
\end{equation}
as the total number of particles that stem from feed-down and 
\begin{equation}
 X_M = \sum_{N_F+1}^{N_M} X_i,
\end{equation}
as the total number of particles that were misidentified (i.e. impurities). $X_0$ is the number of correctly identified primary particles 
that are of interest for the femtoscopy analysis.
\\
The purity $\mathcal{P}$ is the fraction of correctly identified particles, not necessarily primary,
to the total number of particles in the sample (Eq. \ref{eq:App:Purity}). 
\begin{align}\label{eq:App:Purity}
  \mathcal{P}(X) = (X_0+X_F)/X.
\end{align}
The impurity is 
\begin{align}\label{eq:App:Impurity}
  \mathcal{\bar{P}}(X) = X_M/X.
\end{align}
For the later discussion it is beneficial to combine the two definitions and refer to the purity as
\begin{align}\label{eq:App:TotPurity}
  \mathcal{P}(X_i) =
    \begin{cases}
    \mathcal{P}(X)=(X_0+X_F)/X&\text{ for }i\leq N_F,\\
    \mathcal{\bar{P}}(X)=X_M/X&\text{ else.}
    \end{cases}
\end{align}
Another quantity of interest will be the channel fraction $f_i$, which is defined as the fraction of particles originating from the 
$i$-th channel relative to the total number of either correctly identified or misidentified particles:
\begin{align}\label{eq:App:Fractions}
  f(X_i) =
    \begin{cases}
    X_i/(X_0+X_F)&\text{ for }i\leq N_F,\\
    X_i/X_M&\text{ else.}
    \end{cases}
\end{align}
As discussed in the main body of the paper both the purity and the channel fractions can be obtained either from MC simulations or 
MC template fits. The product of the two reads
\begin{equation}\label{eq:App:Pf}
 P(X_i) = \mathcal{P}(X_i)f(X_i) = \frac{X_i}{X}.
\end{equation}
\\
Next we will relate $\mathcal{P}(X_i)$ and $f(X_i)$ to the correlation function between particle pairs, which is defined as 
\begin{equation}
 C(XY) = \frac{N(XY)}{M(XY)},
\end{equation}
where $N$ and $M$ are the yields of an 'XY' particle pair in same and mixed events respectively. Note that this is a raw correlation function 
which is not properly normalized. The normalization is discussed in the main body of the paper, but is irrelevant in the current discussion 
and it will be omitted. Both $N$ and $M$ are yields which can be decomposed into the sum of their ingredients. 
Using the previously discussed notion of different channels of origin
\begin{equation}
 N(XY) = N\left(\sum_{i,j} X_iY_j\right) = \sum_{i,j} N(X_iY_j),
\end{equation}
\begin{equation}
 M(XY) = M\left(\sum_{i,j} X_iY_j\right) = \sum_{i,j} M(X_iY_j).
\end{equation}
Hence the total correlation function becomes:
\begin{align}\label{eq:App:TotCk}
 C(XY) &= \frac{\sum_{i,j} N(X_iY_j)}{M(XY)} =
 \sum_{i,j} \frac{N(X_iY_j)}{M(XY)} \frac{M(X_iY_j)}{M(X_iY_j)}=\\
 &=\sum_{i,j} \underbrace{\frac{N(X_iY_j)}{M(X_iY_j)}}_{C_{i,j}(XY)} \underbrace{\frac{M(X_iY_j)}{M(XY)}}_{\lambda_{i,j}(XY)}=
 \sum_{i,j} \lambda_{i,j}(XY)C_{i,j}(XY),
\end{align}
where $C_{i,j}(XY)$ is the contribution to the total correlation of the $i,j$-th channel of origin of the particles 'X,Y' and 
$\lambda_{i,j}(XY)$ is the corresponding weight coefficient. How to obtain the individual functions $C_{i,j}(XY)$ is 
discussed in the main body of the paper. The weights $\lambda_{i,j}$ can be derived from the purities and channel fractions 
of the particles 'X' and 'Y'. This is possible since $\lambda_{i,j}$ depends only on the mixed event sample 
for which the underlying assumption is that the particles are not correlated. 
In that case the two-particle yield $M(XY)$ can be factorized and according to Eq.~(\ref{eq:App:TotCk}) the $\lambda$ coefficients 
can be expressed as
\begin{equation}
 \lambda_{i,j}(XY)=\frac{M(X_iY_j)}{M(XY)}=\frac{M(X_i)}{M(X)}\frac{M(Y_i)}{M(Y)}=P(X_i)P(Y_i).
\end{equation}
The last step follows directly from Eq.~(\ref{eq:App:Pf}) applied to the mixed event samples of 'X' and 'Y'. 
Eq. \ref{eq:App:Pf} relates $P$ to the known quantities $\mathcal{P}$ and $f$, hence the $\lambda$ coefficients 
can be rewritten as
\begin{equation}
 \lambda_{i,j}(XY)=\mathcal{P}(X_i)f(X_i)\mathcal{P}(Y_j)f(Y_j).
\end{equation}
We would like to point out that due to the definition of $P(X_i)$ the sum of all $\lambda$ parameters is 
automatically normalized to unity.
\newpage
\section{The ALICE Collaboration}
\label{app:collab}

\begingroup
\small
\begin{flushleft}
S.~Acharya\Irefn{org139}\And 
F.T.-.~Acosta\Irefn{org20}\And 
D.~Adamov\'{a}\Irefn{org93}\And 
J.~Adolfsson\Irefn{org80}\And 
M.M.~Aggarwal\Irefn{org98}\And 
G.~Aglieri Rinella\Irefn{org34}\And 
M.~Agnello\Irefn{org31}\And 
N.~Agrawal\Irefn{org48}\And 
Z.~Ahammed\Irefn{org139}\And 
S.U.~Ahn\Irefn{org76}\And 
S.~Aiola\Irefn{org144}
% ...
\And
\renewcommand\labelenumi{\textsuperscript{\theenumi}~}

\section*{Affiliation notes}
\renewcommand\theenumi{\roman{enumi}}
\begin{Authlist}
\item \Adef{org*}Deceased
\item \Adef{orgI}Dipartimento DET del Politecnico di Torino, Turin, Italy
\item \Adef{orgII}M.V. Lomonosov Moscow State University, D.V. Skobeltsyn Institute of Nuclear, Physics, Moscow, Russia
\item \Adef{orgIII}Department of Applied Physics, Aligarh Muslim University, Aligarh, India
\item \Adef{orgIV}Institute of Theoretical Physics, University of Wroclaw, Poland
\end{Authlist}

\section*{Collaboration Institutes}
\renewcommand\theenumi{\arabic{enumi}~}
\begin{Authlist}
\item \Idef{org1}A.I. Alikhanyan National Science Laboratory (Yerevan Physics Institute) Foundation, Yerevan, Armenia
\item \Idef{org2}Bogolyubov Institute for Theoretical Physics, National Academy of Sciences of Ukraine, Kiev, Ukraine
\item \Idef{org3}Bose Institute, Department of Physics  and Centre for Astroparticle Physics and Space Science (CAPSS), Kolkata, India
\item \Idef{org4}Budker Institute for Nuclear Physics, Novosibirsk, Russia
\item \Idef{org5}California Polytechnic State University, San Luis Obispo, California, United States
\item \Idef{org6}Central China Normal University, Wuhan, China
\item \Idef{org7}Centre de Calcul de l'IN2P3, Villeurbanne, Lyon, France
\item \Idef{org8}Centro de Aplicaciones Tecnol\'{o}gicas y Desarrollo Nuclear (CEADEN), Havana, Cuba
\item \Idef{org9}Centro de Investigaci\'{o}n y de Estudios Avanzados (CINVESTAV), Mexico City and M\'{e}rida, Mexico
\item \Idef{org10}Centro Fermi - Museo Storico della Fisica e Centro Studi e Ricerche ``Enrico Fermi', Rome, Italy
\item \Idef{org11}Chicago State University, Chicago, Illinois, United States
\item \Idef{org12}China Institute of Atomic Energy, Beijing, China
\item \Idef{org13}Chonbuk National University, Jeonju, Republic of Korea
\item \Idef{org14}Comenius University Bratislava, Faculty of Mathematics, Physics and Informatics, Bratislava, Slovakia
\item \Idef{org15}COMSATS Institute of Information Technology (CIIT), Islamabad, Pakistan
\item \Idef{org16}Creighton University, Omaha, Nebraska, United States
\item \Idef{org17}Department of Physics, Aligarh Muslim University, Aligarh, India
\item \Idef{org18}Department of Physics, Pusan National University, Pusan, Republic of Korea
\item \Idef{org19}Department of Physics, Sejong University, Seoul, Republic of Korea
\item \Idef{org20}Department of Physics, University of California, Berkeley, California, United States
\item \Idef{org21}Department of Physics, University of Oslo, Oslo, Norway
\item \Idef{org22}Department of Physics and Technology, University of Bergen, Bergen, Norway
\item \Idef{org23}Dipartimento di Fisica dell'Universit\`{a} 'La Sapienza' and Sezione INFN, Rome, Italy
\item \Idef{org24}Dipartimento di Fisica dell'Universit\`{a} and Sezione INFN, Cagliari, Italy
\item \Idef{org25}Dipartimento di Fisica dell'Universit\`{a} and Sezione INFN, Trieste, Italy
\item \Idef{org26}Dipartimento di Fisica dell'Universit\`{a} and Sezione INFN, Turin, Italy
\item \Idef{org27}Dipartimento di Fisica e Astronomia dell'Universit\`{a} and Sezione INFN, Bologna, Italy
\item \Idef{org28}Dipartimento di Fisica e Astronomia dell'Universit\`{a} and Sezione INFN, Catania, Italy
\item \Idef{org29}Dipartimento di Fisica e Astronomia dell'Universit\`{a} and Sezione INFN, Padova, Italy
\item \Idef{org30}Dipartimento di Fisica `E.R.~Caianiello' dell'Universit\`{a} and Gruppo Collegato INFN, Salerno, Italy
\item \Idef{org31}Dipartimento DISAT del Politecnico and Sezione INFN, Turin, Italy
\item \Idef{org32}Dipartimento di Scienze e Innovazione Tecnologica dell'Universit\`{a} del Piemonte Orientale and INFN Sezione di Torino, Alessandria, Italy
\item \Idef{org33}Dipartimento Interateneo di Fisica `M.~Merlin' and Sezione INFN, Bari, Italy
\item \Idef{org34}European Organization for Nuclear Research (CERN), Geneva, Switzerland
\item \Idef{org35}Faculty of Electrical Engineering, Mechanical Engineering and Naval Architecture, University of Split, Split, Croatia
\item \Idef{org36}Faculty of Engineering and Science, Western Norway University of Applied Sciences, Bergen, Norway
\item \Idef{org37}Faculty of Nuclear Sciences and Physical Engineering, Czech Technical University in Prague, Prague, Czech Republic
\item \Idef{org38}Faculty of Science, P.J.~\v{S}af\'{a}rik University, Ko\v{s}ice, Slovakia
\item \Idef{org39}Frankfurt Institute for Advanced Studies, Johann Wolfgang Goethe-Universit\"{a}t Frankfurt, Frankfurt, Germany
\item \Idef{org40}Gangneung-Wonju National University, Gangneung, Republic of Korea
\item \Idef{org41}Gauhati University, Department of Physics, Guwahati, India
\item \Idef{org42}Helmholtz-Institut f\"{u}r Strahlen- und Kernphysik, Rheinische Friedrich-Wilhelms-Universit\"{a}t Bonn, Bonn, Germany
\item \Idef{org43}Helsinki Institute of Physics (HIP), Helsinki, Finland
\item \Idef{org44}High Energy Physics Group,  Universidad Aut\'{o}noma de Puebla, Puebla, Mexico
\item \Idef{org45}Hiroshima University, Hiroshima, Japan
\item \Idef{org46}Hochschule Worms, Zentrum  f\"{u}r Technologietransfer und Telekommunikation (ZTT), Worms, Germany
\item \Idef{org47}Horia Hulubei National Institute of Physics and Nuclear Engineering, Bucharest, Romania
\item \Idef{org48}Indian Institute of Technology Bombay (IIT), Mumbai, India
\item \Idef{org49}Indian Institute of Technology Indore, Indore, India
\item \Idef{org50}Indonesian Institute of Sciences, Jakarta, Indonesia
\item \Idef{org51}INFN, Laboratori Nazionali di Frascati, Frascati, Italy
\item \Idef{org52}INFN, Sezione di Bari, Bari, Italy
\item \Idef{org53}INFN, Sezione di Bologna, Bologna, Italy
\item \Idef{org54}INFN, Sezione di Cagliari, Cagliari, Italy
\item \Idef{org55}INFN, Sezione di Catania, Catania, Italy
\item \Idef{org56}INFN, Sezione di Padova, Padova, Italy
\item \Idef{org57}INFN, Sezione di Roma, Rome, Italy
\item \Idef{org58}INFN, Sezione di Torino, Turin, Italy
\item \Idef{org59}INFN, Sezione di Trieste, Trieste, Italy
\item \Idef{org60}Inha University, Incheon, Republic of Korea
\item \Idef{org61}Institut de Physique Nucl\'{e}aire d'Orsay (IPNO), Institut National de Physique Nucl\'{e}aire et de Physique des Particules (IN2P3/CNRS), Universit\'{e} de Paris-Sud, Universit\'{e} Paris-Saclay, Orsay, France
\item \Idef{org62}Institute for Nuclear Research, Academy of Sciences, Moscow, Russia
\item \Idef{org63}Institute for Subatomic Physics, Utrecht University/Nikhef, Utrecht, Netherlands
\item \Idef{org64}Institute for Theoretical and Experimental Physics, Moscow, Russia
\item \Idef{org65}Institute of Experimental Physics, Slovak Academy of Sciences, Ko\v{s}ice, Slovakia
\item \Idef{org66}Institute of Physics, Homi Bhabha National Institute, Bhubaneswar, India
\item \Idef{org67}Institute of Physics of the Czech Academy of Sciences, Prague, Czech Republic
\item \Idef{org68}Institute of Space Science (ISS), Bucharest, Romania
\item \Idef{org69}Institut f\"{u}r Kernphysik, Johann Wolfgang Goethe-Universit\"{a}t Frankfurt, Frankfurt, Germany
\item \Idef{org70}Instituto de Ciencias Nucleares, Universidad Nacional Aut\'{o}noma de M\'{e}xico, Mexico City, Mexico
\item \Idef{org71}Instituto de F\'{i}sica, Universidade Federal do Rio Grande do Sul (UFRGS), Porto Alegre, Brazil
\item \Idef{org72}Instituto de F\'{\i}sica, Universidad Nacional Aut\'{o}noma de M\'{e}xico, Mexico City, Mexico
\item \Idef{org73}iThemba LABS, National Research Foundation, Somerset West, South Africa
\item \Idef{org74}Johann-Wolfgang-Goethe Universit\"{a}t Frankfurt Institut f\"{u}r Informatik, Fachbereich Informatik und Mathematik, Frankfurt, Germany
\item \Idef{org75}Joint Institute for Nuclear Research (JINR), Dubna, Russia
\item \Idef{org76}Korea Institute of Science and Technology Information, Daejeon, Republic of Korea
\item \Idef{org77}KTO Karatay University, Konya, Turkey
\item \Idef{org78}Laboratoire de Physique Subatomique et de Cosmologie, Universit\'{e} Grenoble-Alpes, CNRS-IN2P3, Grenoble, France
\item \Idef{org79}Lawrence Berkeley National Laboratory, Berkeley, California, United States
\item \Idef{org80}Lund University Department of Physics, Division of Particle Physics, Lund, Sweden
\item \Idef{org81}Nagasaki Institute of Applied Science, Nagasaki, Japan
\item \Idef{org82}Nara Women{'}s University (NWU), Nara, Japan
\item \Idef{org83}National and Kapodistrian University of Athens, School of Science, Department of Physics , Athens, Greece
\item \Idef{org84}National Centre for Nuclear Research, Warsaw, Poland
\item \Idef{org85}National Institute of Science Education and Research, Homi Bhabha National Institute, Jatni, India
\item \Idef{org86}National Nuclear Research Center, Baku, Azerbaijan
\item \Idef{org87}National Research Centre Kurchatov Institute, Moscow, Russia
\item \Idef{org88}Niels Bohr Institute, University of Copenhagen, Copenhagen, Denmark
\item \Idef{org89}Nikhef, National institute for subatomic physics, Amsterdam, Netherlands
\item \Idef{org90}NRC Kurchatov Institute IHEP, Protvino, Russia
\item \Idef{org91}NRNU Moscow Engineering Physics Institute, Moscow, Russia
\item \Idef{org92}Nuclear Physics Group, STFC Daresbury Laboratory, Daresbury, United Kingdom
\item \Idef{org93}Nuclear Physics Institute of the Czech Academy of Sciences, \v{R}e\v{z} u Prahy, Czech Republic
\item \Idef{org94}Oak Ridge National Laboratory, Oak Ridge, Tennessee, United States
\item \Idef{org95}Ohio State University, Columbus, Ohio, United States
\item \Idef{org96}Petersburg Nuclear Physics Institute, Gatchina, Russia
\item \Idef{org97}Physics department, Faculty of science, University of Zagreb, Zagreb, Croatia
\item \Idef{org98}Physics Department, Panjab University, Chandigarh, India
\item \Idef{org99}Physics Department, University of Jammu, Jammu, India
\item \Idef{org100}Physics Department, University of Rajasthan, Jaipur, India
\item \Idef{org101}Physikalisches Institut, Eberhard-Karls-Universit\"{a}t T\"{u}bingen, T\"{u}bingen, Germany
\item \Idef{org102}Physikalisches Institut, Ruprecht-Karls-Universit\"{a}t Heidelberg, Heidelberg, Germany
\item \Idef{org103}Physik Department, Technische Universit\"{a}t M\"{u}nchen, Munich, Germany
\item \Idef{org104}Research Division and ExtreMe Matter Institute EMMI, GSI Helmholtzzentrum f\"ur Schwerionenforschung GmbH, Darmstadt, Germany
\item \Idef{org105}Rudjer Bo\v{s}kovi\'{c} Institute, Zagreb, Croatia
\item \Idef{org106}Russian Federal Nuclear Center (VNIIEF), Sarov, Russia
\item \Idef{org107}Saha Institute of Nuclear Physics, Homi Bhabha National Institute, Kolkata, India
\item \Idef{org108}School of Physics and Astronomy, University of Birmingham, Birmingham, United Kingdom
\item \Idef{org109}Secci\'{o}n F\'{\i}sica, Departamento de Ciencias, Pontificia Universidad Cat\'{o}lica del Per\'{u}, Lima, Peru
\item \Idef{org110}Shanghai Institute of Applied Physics, Shanghai, China
\item \Idef{org111}St. Petersburg State University, St. Petersburg, Russia
\item \Idef{org112}Stefan Meyer Institut f\"{u}r Subatomare Physik (SMI), Vienna, Austria
\item \Idef{org113}SUBATECH, IMT Atlantique, Universit\'{e} de Nantes, CNRS-IN2P3, Nantes, France
\item \Idef{org114}Suranaree University of Technology, Nakhon Ratchasima, Thailand
\item \Idef{org115}Technical University of Ko\v{s}ice, Ko\v{s}ice, Slovakia
\item \Idef{org116}Technische Universit\"{a}t M\"{u}nchen, Excellence Cluster 'Universe', Munich, Germany
\item \Idef{org117}The Henryk Niewodniczanski Institute of Nuclear Physics, Polish Academy of Sciences, Cracow, Poland
\item \Idef{org118}The University of Texas at Austin, Austin, Texas, United States
\item \Idef{org119}Universidad Aut\'{o}noma de Sinaloa, Culiac\'{a}n, Mexico
\item \Idef{org120}Universidade de S\~{a}o Paulo (USP), S\~{a}o Paulo, Brazil
\item \Idef{org121}Universidade Estadual de Campinas (UNICAMP), Campinas, Brazil
\item \Idef{org122}Universidade Federal do ABC, Santo Andre, Brazil
\item \Idef{org123}University College of Southeast Norway, Tonsberg, Norway
\item \Idef{org124}University of Cape Town, Cape Town, South Africa
\item \Idef{org125}University of Houston, Houston, Texas, United States
\item \Idef{org126}University of Jyv\"{a}skyl\"{a}, Jyv\"{a}skyl\"{a}, Finland
\item \Idef{org127}University of Liverpool, Liverpool, United Kingdom
\item \Idef{org128}University of Tennessee, Knoxville, Tennessee, United States
\item \Idef{org129}University of the Witwatersrand, Johannesburg, South Africa
\item \Idef{org130}University of Tokyo, Tokyo, Japan
\item \Idef{org131}University of Tsukuba, Tsukuba, Japan
\item \Idef{org132}Universit\'{e} Clermont Auvergne, CNRS/IN2P3, LPC, Clermont-Ferrand, France
\item \Idef{org133}Universit\'{e} de Lyon, Universit\'{e} Lyon 1, CNRS/IN2P3, IPN-Lyon, Villeurbanne, Lyon, France
\item \Idef{org134}Universit\'{e} de Strasbourg, CNRS, IPHC UMR 7178, F-67000 Strasbourg, France, Strasbourg, France
\item \Idef{org135} Universit\'{e} Paris-Saclay Centre d¿\'Etudes de Saclay (CEA), IRFU, Department de Physique Nucl\'{e}aire (DPhN), Saclay, France
\item \Idef{org136}Universit\`{a} degli Studi di Foggia, Foggia, Italy
\item \Idef{org137}Universit\`{a} degli Studi di Pavia, Pavia, Italy
\item \Idef{org138}Universit\`{a} di Brescia, Brescia, Italy
\item \Idef{org139}Variable Energy Cyclotron Centre, Homi Bhabha National Institute, Kolkata, India
\item \Idef{org140}Warsaw University of Technology, Warsaw, Poland
\item \Idef{org141}Wayne State University, Detroit, Michigan, United States
\item \Idef{org142}Westf\"{a}lische Wilhelms-Universit\"{a}t M\"{u}nster, Institut f\"{u}r Kernphysik, M\"{u}nster, Germany
\item \Idef{org143}Wigner Research Centre for Physics, Hungarian Academy of Sciences, Budapest, Hungary
\item \Idef{org144}Yale University, New Haven, Connecticut, United States
\item \Idef{org145}Yonsei University, Seoul, Republic of Korea
\end{Authlist}
\endgroup
\end{document}